\documentclass[12pt]{article}
\usepackage[utf8]{inputenc}
\usepackage{mathtools}
\usepackage{natbib}
\usepackage{threeparttable}
\usepackage{array}
\usepackage{lscape}
\usepackage{adjustbox}

\usepackage{amssymb,amsmath, amsthm}
\usepackage{bbm,comment}
\usepackage{setspace}
\usepackage[margin=1.25in]{geometry}

\usepackage{color}

\newcommand{\K}{\mathcal{K}}
\newcommand{\W}{\mathcal{W}}

\DeclareMathOperator*{\argmax}{arg\,max}

\def\p{\partial}
\def\ps{p^{*}}
\def\Re{\mathbb{R}}
\def\mB{\mathcal{B}}

\def\mN{\mathcal{N}}
\def\Prob{\mathbb{P}}

\def\rt#1{\sqrt{#1}\,}

\def\nano{\scriptscriptstyle}

\newcommand\ca[1]{{\cal{#1}}}
\newcommand\lo[1]{_{\nano #1}}
\newcommand*\diff{\mathop{}\!\mathrm{d}}

\def\E{{\rm E}\,}

\newcommand{\SEES}{\text{SEES}}

\newtheorem{theorem}{Theorem}
\newtheorem{corollary}{Corollary}[theorem]
\newtheorem{lemma}[theorem]{Lemma}
\newtheorem{proposition}{Proposition}

\theoremstyle{definition}
\newtheorem{assump}{{\bf Assumption}}
\newtheorem{assumpB}{{\bf Assumption}}

\theoremstyle{remark}
\newtheorem{rmk}{{\bf Remark}}

\title{Efficient Estimation of Structural Models via Sieves\thanks{First version: February 2022. \textit{Contact Information}: Luo: Department of Economics, University of Toronto, Max Gluskin House, 150 St. George St, Toronto, ON M5S 3G7, Canada (email: yao.luo@utoronto.ca); Sang: Department of Statistics and Actuarial Science, University of Waterloo, 200 University Avenue West, Waterloo, ON, Canada, N2L 3G1, Canada (email: psang@uwaterloo.ca). We are grateful to Victor Aguirregabiria, Xiaohong Chen, Paul Grieco, Marc Rysman and seminar participants at the Canadian Economic Association, ICSA-Canada, Asian Meeting of the Econometric Society, CESG 2023, University of Toronto and Western University for helpful discussions and comments. Luo acknowledges the Social Sciences and Humanities Research Council of Canada for research support. All errors are our own.
}}

\author{Yao Luo$^\dagger$ \and Peijun Sang$^\ddag$}
\date{%
    $^\dagger$University of Toronto\\%
    $^\ddag$University of Waterloo\\[2ex]%
    September 16, 2024
}

\begin{document}

\maketitle


\begin{abstract}
We propose a class of sieve-based efficient estimators for structural models (SEES), which approximate the solution using a linear combination of basis functions and impose equilibrium conditions as a penalty to determine the best-fitting coefficients. Our estimators avoid the need to repeatedly solve the model, apply to a broad class of models, and are consistent, asymptotically normal, and asymptotically efficient. Moreover, they solve \textit{unconstrained} optimization problems with fewer unknowns and offer convenient standard error calculations. As an illustration, we apply our method to an entry game between Walmart and Kmart.
\end{abstract}



\textbf{Keywords:} Efficient Estimation, Sieves, Empirical Games, Joint Algorithm, Nested Algorithm

\newpage

\section{Introduction}



A structural model builds on economic theory and describes how a set of endogenous variables are related to a set of explanatory variables. This relation is often in the form of an implicit function. In particular, it generates endogenous function $p$ determined by an equation system: 
\begin{equation}
p = \Psi(p,\theta) ,
\label{eq-structure}
\end{equation}
where $\theta$ is the parameter of interest and $\Psi$ is a representation of the structural model.\footnote{In discrete choice models, the parameter captures consumer preferences and the observable is consumer choice; in auctions, the parameter captures the value distribution and the observable is the bid distribution; in dynamic models, the parameter describes the agent's intertemporal tradeoff and the observable is intertemporal choice.} While $\Psi$ is explicit, solving for $p$ could be difficult or costly. Such computational burden limits the use of standard estimators. For instance, the maximum likelihood estimator (MLE) repeatedly guesses $\theta$ and evaluates data likelihood using the solution of \eqref{eq-structure}, $p^*(\theta)$. However, finding the solution can be computationally intensive and is often hindered by a lack of robust algorithms, particularly in empirical games.






We introduce a new class of estimators: sieve-based efficient estimators for structural models (\SEES). Our approach applies to a broad class of models, including empirical games, and avoids solving the model. Our \SEES\ is motivated by two popular approaches to infinite-dimensional optimization problems: approximation and penalization. See \citet{shen1997methods}, \citet{shen1998method}, and \citet{chen2007large}. Because the likelihood function $\ell(p^*,\theta; \text{data})$ involves an unknown function $p^*$, maximizing the likelihood with respect to $p^*$ and $\theta$ may lead to an asymptotically inefficient estimator for the parameter, and the resulting estimator may not necessarily be close to the solution of \eqref{eq-structure}. To address these issues, prior studies utilize sieves that are less complex but dense to approximate the original function space, and regularization that assumes smoothness of this function. 

In this paper, we estimate structural models by approximating the solution to avoid solving the model and regularizing with the equilibrium conditions that are built into the model itself. By combining the data fitting and model fitting criteria, we formulate our penalized log-likelihood criterion,
\[
\underbrace{\ell(\beta,\theta;\text{data})}_{\text{data likelihood}} - \underbrace{\omega \times \rho(\beta,\theta)}_{\text{penalization}},
\]
where $\ell$ and $\rho$ measure the data fitting and model fitting, respectively.\footnote{While our idea extends to other types of estimators, we focus on likelihood-based ones here.} Moreover, $\beta \in \Re^K$ and $\omega \in \Re_+$ govern the approximation and the weighting, respectively. Instead of imposing stronger smoothness assumptions than typically implied by theory, our approach relies solely on the model to regularize the sieve approximation. The smoothing parameter $\omega$ explicitly captures the weighting of the data likelihood and the equilibrium condition, and the dimension of the approximation parameter $\beta$, denoted by $K$, balances computational cost and solution accuracy.

Allowing these tuning parameters to diverge at appropriate rates, the proposed parameter estimator of $\theta$ is consistent, asymptotically normal, and asymptotically efficient. Intuitively, by gradually updating the smoothing parameter, we shift the weight from the data to the equilibrium condition. At the minimum, a preliminary nonparametric estimate of $p$ (by letting $\omega =0$) constitutes a good starting value but is subject to issues with nonparametric estimates. When the smoothing parameter increases, more weight is given to the equilibrium condition. By forcing model restrictions more strongly, the estimator converges to the MLE.

We prescribe several algorithms to implement \SEES. The first is a \textit{joint} algorithm that finds the combination of sieve approximations and model parameters that best explains the data and satisfies the equilibrium conditions. That is, we maximize the penalized log-likelihood function with respect to $(\beta,\theta)$. The second is a \textit{nested} algorithm that consists of two main parts. First, for each model parameter $\theta$, we find the sieve approximation of $p$ that best explains the data and satisfies the equilibrium conditions. Second, based on the approximated solution, we find the model parameter that best fits the data. While the joint algorithm is attractive because it results in a single-level optimization problem, the nested algorithm is quite intuitive, resembling MLE. 


Our estimator allows for discrete and continuous state/heterogeneity in the model to be estimated. The standard practice of discretizing continuous state variables or covariates leads to efficiency loss. Under mild regularity conditions, we show that our estimator has the same asymptotic distribution as MLE in both cases. To our knowledge, we are the first to combine approximation and penalization in estimating structural models. While some studies have adopted approximation approaches, none combines them with penalization. Another important advantage of our method is that it produces standard errors in the same way as the standard MLE using the Fisher information matrix, which is of considerable convenience in empirical work. As a side product, we also derive a similar approach for the mathematical program with equilibrium constraints (MPEC) estimators, which provides a faster alternative than the bootstrapping method previously proposed by \citet{su2012constrained}.


We acknowledge several limitations inherent in our methodology. First, our sieve-based approach presupposes the smoothness of the solution within the state variables or covariates, leaving the treatment of discontinuities as a subject for future investigation. Second, our approach provides a robust solution that works with minimal assumptions on the solution, which is particularly valuable in models with unfavorable or unknown properties. However, it may not always be the most expedient choice in scenarios where the solution exhibits favorable properties, such as contraction mappings. For instance, as demonstrated in the simple model outlined in Section \ref{sec:example}, it exhibits a relatively slower performance when compared to a nested-fixed point algorithm. Throughout this paper, we refrain from comparing computational time across different estimators, as it is often model-specific and, hence, more relevant in richer empirical models.

The remainder of the paper is organized as follows. Section \ref{sec:example} explains the idea using a simple example. Section \ref{profile} proposes the class of sieve-based efficient estimators for structural models and derives its asymptotic properties. Section \ref{sec:app} demonstrates the performance of our estimators in estimating an empirical game. Section \ref{sec:conc} concludes. The Appendix contains all omitted proofs and details. 

\section{A Motivating Example} \label{sec:example}

Our new method differs from existing methods by how we leverage data and model restrictions. We now compare it with popular existing methods, such as maximum likelihood estimation (MLE), two-step approaches, and nested pseudo-likelihood (NPL), through a motivating example. 

Consider a monopolist $j$, facing logit demand, sells a product at a price $P_j$. That is, consumer $i$ gets utility of 
\[
u_{ij} = \xi_j - \alpha P_j + \epsilon_{ij} , 
\]
where $\xi_j$ is \textit{continuous} product quality, $\alpha$ is the price coefficient, and $\epsilon_{ij}$ represents the standard Type 1 extreme value (T1EV) taste shock. The firm's profit maximization problem is 
\[
\max_{P_j} \quad 
(\underbrace{P_j - c_j}_{\text{profit margin}}) \times 
\underbrace{\frac{\exp(\xi_j - \alpha P_j)}{1+\exp(\xi_j - \alpha P_j)}}_{\text{market share}} ,
\]
where $c_j$ represents the constant marginal cost. The optimal price is determined by the FOC, 
\[
\alpha(P_j - c_j) = 1+\exp(\xi_j - \alpha P_j),
\]
where $P_j$ appears both inside and outside an exponential function. As a result, the mapping from the parameters to the optimal price is implicit.



\subsection{Structural Estimation}

For simplicity, we focus on estimating the parameter $\theta$ that governs consumer preferences over product feature $x_j \in \mathbb{R}$ using observed prices. Specifically, we treat it as known that $c_j = 0,~\alpha = 1, ~\xi_j = \log x_j + \log \theta + 1$, where ``1'' is quality normalization for simplicity. Appendix \ref{lambertprice} shows that the optimal price satisfies 
\begin{equation}\label{Plogit}
y_j^* = p(x_j; \theta), 
\end{equation}
where $y_j^* = P_j^* -1 $ represents the normalized price and $p(x_j; \theta)$ is defined by 
\[
p(x_j; \theta) e^{p(x_j; \theta)} = \theta x_j \quad \text{ or } \quad
p(x_j; \theta) = \theta x_j e^{-p(x_j; \theta)} ,
\]
the first of which has the standard form of the Lambert W function\footnote{The Lambert W function $W(x)$ is defined by $W(x) e^{W(x)} = x$.} and the second of which has the same form as Equation \eqref{eq-structure}. We denote this function as $p(\cdot; \theta)$ to indicate its dependence on the parameter. 

Consider a data generating process (DGP) that is a noisy measurement of the optimal price $y_j = y_j^* + e_j$, where $e_j$'s are measurement errors that are i.i.d. draws from the standard normal distribution. Therefore, the observed (normalized) price $y_j$ is from the standard normal distribution with a location $p(x_j;\theta_0)$: 
\[
y_j \sim \mathcal{N}(p(x_j;\theta_0), 1), \text{ where } j=1,\ldots, N .
\]
The data contain the product characteristics $x_j$ and the prices $P_j= P_j^* + e_j$ (equivalently, the normalized prices $y_j$). The parameter of interest is $\theta$.


\vspace{.5cm}
\noindent \textbf{Maximum Likelihood Estimation: } The standard MLE solves the following problem 
\[
\max_{\theta} \quad 
\sum_{j=1}^N \log \phi(y_j - p(x_j;\theta)),
\]
where $\phi(\cdot)$ represents the density function of the standard normal distribution. Because $p(\cdot;\theta)$ is implicitly defined, this estimator is computationally costly. For each trial of $\theta$, we need to find $p(x_j;\theta)$ for each data point $x_j$. The number of equations that need to be solved equals the sample size multiplied by the number of likelihood function evaluations. 

Despite its asymptotic efficiency, the standard MLE requires solving the model for each parameter and thus solution algorithms that are sufficiently efficient and robust. When a contraction mapping solution for the model is available, it is often referred to as the nested fixed point algorithm (NFXP). See, e.g., \citet{rust1987optimal} in dynamic discrete choice models and \citet{berry1995automobile} in demand models.


\vspace{.5cm}
\noindent \textbf{A Two-Step Approach: }
We can ``invert the FOC'' and obtain a representation of the ``unknown'' in terms of the optimal prices: 
\[
 \theta = \frac{p(x_j; \theta) e^{p(x_j; \theta)} }{x}  ,
\]
where the normalized price $y_j^*= p(x_j; \theta)$ is unobserved. In principle, this FOC inversion allows estimating the parameter using the optimal price in any market.

Due to measurement errors, the rewritten FOC suggests a simple two-step approach that avoids solving the model repeatedly in estimation. In the first step, we consider $y_j^* = p(x_j; \theta_0)$ and estimate the optimal price as a function of the covariate. Although the true parameter $\theta_0$, the endogenous variable $p$ and thus the RHS $p(x_j; \theta_0)$ are all unobserved, we can estimate the LHS $y_j^*(x_j)$ nonparametrically using the observed price and covariate pairs $\{x_j, y_j\}_{j=1}^N$. In particular, we run a nonparametric regression\footnote{We apply kernel regression using the optimal bandwidth estimated by cross-validation, as provided on Professor Yingying Dong's webpage: http://yingyingdong.com.}, 
\[
y_j = y_j^*(x_j) +e ,
\]
and obtain an estimate of the normalized price $\widehat{y_j^*(x_j)}$. 
In the second step, we have a simple plug-in estimator, 
\[
\widehat{\theta } = \text{median}\left\{\frac{\widehat{y_j^*(x_j)} \times \exp{\widehat{y_j^*(x_j)} } }{x_j}: j = 1, \ldots, N\right\}.
\]

Two-step approaches avoid repeatedly solving the economic model at the expense of efficiency. In the first step, the analyst obtains a nonparametric estimate of the endogenous variable $\widehat{p}$. In the second step, the estimate is obtained from $\widehat{p} = \Psi(\widehat{p}, \theta)$ in various ways. In auction models, \citet{guerre2000optimal} use the estimated bid distribution to construct pseudo values, which are then used to estimate the underlying value distribution. In dynamic discrete choice models, the conditional choice probability (CCP) approach of \citet{hotz1993conditional} plugs the estimated CCPs into the optimal decision rules. In dynamic games, one can obtain a nonlinear least squares estimate of $\theta$ by replacing $p$ with the estimated CCP $\widehat{p}$ in the function; see \citet{pesendorfer2008asymptotic}.


\vspace{.5cm}
\noindent \textbf{Nested Pseudo-Likelihood Algorithm: }
In each iteration, the NPL algorithm solves the following problem:
\[
\max_{\theta} \quad 
\sum_{j=1}^N \log \phi(y_j -\theta x_j \exp(-\widehat{p}_j^\tau)),
\]
where $\widehat{p}_j^\tau$ represents some estimate of the optimal price in market $j$. Denote the solution as $\widehat{\theta}^\tau$. We can then update the price estimates $\widehat{p}_j^{\tau+1} = \widehat{\theta}^\tau x_j \exp(-\widehat{p}_j^\tau)$. We iterate the process till the parameter estimate converges.


Given some estimates $\widehat{\theta}$ and $\widehat{p}$, the NPL algorithm obtains new estimates of the choice probabilities by applying the mapping $\tilde{p} = \Psi(\widehat{p},\widehat{\theta})$ and then updates the parameter estimate by maximizing the pseudo-likelihood function $\ell(\tilde{p}, \theta)$.


\vspace{.5cm}
\noindent \textbf{A Sieve-Based Efficient Estimator: }
In this paper, we propose a method that obviates solving the model repeatedly. In particular, we approximate the ``solution'' function $p^{\star}(\cdot)$ by B-spline basis functions: 
\[
p^\beta(\cdot) = \sum_{k=1}^K \beta_k s_k(\cdot),
\] 
where $s_k(\cdot)$ is a cubic spline basis function, and $K$ denotes the number of basis functions. Our 
sieve-based estimator of $\theta$ maximizes the likelihood 
\[
\sum_{j=1}^N \log \left\{\phi\left(y_j - p^{\widehat{\beta}(\theta,\omega)}(x_j)\right)\right\} ,
\]
where $\widehat{\beta}(\theta,\omega)$ is defined by
\[
\arg\max_{\beta \in \Re^K} \quad 
\sum_{j=1}^N \log \left\{\phi(y_j - p^\beta(x_j)) \right\}
- \underbrace{\omega \int_{\mathcal{X}}  \left[p^\beta(x) e^{p^\beta(x)} - \theta x\right]^2  dx}_{\text{penalization}},
\]
where $\omega >0$.
Because $p^\beta(\cdot)$ is an approximation, the second term penalizes the likelihood by the amount of deviation by definition of the Lambert W function.

\subsubsection*{Discussion}

We now compare the above-mentioned estimators. First, \SEES\ and MLE are asymptotically equivalent and almost identical in finite samples. However, NFXP algorithms may converge slowly, and such mappings may not even exist in important models. For instance, empirical games, such as asymmetric auctions and dynamic games, are notoriously difficult to solve, making MLE difficult to apply. In contrast, we avoid solving the model repeatedly by approximating the solution flexibly. 

Second, two-step approaches are limited by the first-step nonparametric estimation of the endogenous variable and may suffer from the ``curse of dimensionality'' when $x$ has multiple dimensions \citep{stone1980optimal}. As a result, the finite-sample estimation error can be substantial. In contrast, our approximated solution avoids this issue, as its final version relies almost entirely on the model.

Third, our estimator is also related to the nested pseudo-likelihood algorithm proposed by \citet{aguirregabiria2002swapping,aguirregabiria2007sequential}. Exploiting the unique feature of dynamic discrete choice models that the Jacobian matrix of $\Psi_\theta$ is always zero, their iterative refinement converges to MLE. However, it requires some discretion in applying it to empirical games. See \citet{pesendorfer2010sequential}. Both algorithms bridge the gap between the standard MLE and two-step methods, and are asymptotically equivalent to MLE. However, they are based on very different ideas. Our estimator is flexible to accommodate different estimation algorithms, including one that resembles NPL, and robust in applications to various models, including empirical games. 

To our knowledge, we are the first to combine approximation and penalization in estimating structural models. While some studies have adopted approximation approaches, none combines them with penalization. For instance, \citet{keane1994solution,keane1997career} use sieves to approximate solutions in dynamic structural models, combining approximation and NFXP. In estimating dynamic games, \citet{sweeting2013dynamic} uses parametric approximations to the value function, combining approximation and NPL. Most related, \cite{barwick2015costs} approximates the value function using sieves and imposes the Bellman equation as an equilibrium constraint. 

Another related algorithm is MPEC, which is an alternative computational algorithm to MLE. See, e.g., \citet{su2012constrained} and \citet{dube2012improving}.\footnote{Several papers have compared MPEC with the original estimators for various models. For a comparison of NFXP and MPEC, see, e.g., \citet{lee2015computationally} for the \citet{berry1995automobile} model and \citet{iskhakov2016comment} for dynamic models.} It avoids solving the model repeatedly by augmenting the unknown to $(\theta,p)$ and imposing the equilibrium condition as a constraint: 
\begin{align*}
\max_{\theta,p} & \quad
\ell(p,\theta;\text{data})  \\ 
 \text{subject to}  &  \quad p = \Psi(p,\theta) .
\end{align*}

We will show that our \SEES's dual problem is a natural extension of MPEC in discrete state settings. Our \SEES\ nests MPEC as a limiting case when the number of basis functions is the same as the dimension of $p$ and the regularization parameter equals infinity. MPEC forms the Lagrangian function using Lagrange multipliers $\Lambda$ that are of the same size as $p$: $\max_{\theta,p,\Lambda}
\ell(p,\theta;\text{data})   - \Lambda^\prime (p - \Psi(p,\theta))$. As a result, it solves $2 \times \text{dim}(p)+\text{dim}(\theta)$ equations in the same number of unknowns. Our \SEES\ approximates $p$ by $\beta$ and introduces a scalar regularization parameter $\omega$, which reduces the problem to an \textit{unconstrained} optimization problem with $K +1+\text{dim}(\theta) $ unknowns. Therefore, \SEES\ is computationally convenient because $K+1 \ll 2 \times \text{dim}(p)$.




\subsection{Simulation Evidence}

Consider $x_j \sim \text{Uniform}[0,\overline x]$ and $\theta_0 = 1$. For MLE, we use the bisection method to solve for the Lambert W function. We omit MPEC here for two reasons. First, we focus on the statistical properties rather than the computational time of the estimators. Second, MPEC is an alternative computational algorithm to MLE with identical statistical properties. For the two-step approach, we use local linear kernel regression and apply the optimal bandwidth chosen via cross validation. For the proposed method, we use the cubic spline explained in \citet{luo2018structural} and let $K=6$; the choice of $\omega$ follows the method that we propose later. We also provide the analytic gradient of the outer loop and the analytic gradient and Hessian of the inner loop maximization problem; see Appendix \ref{JHprice}. 

\begin{figure}[ht]
\caption{Compare MLE, NPL, and \SEES} \label{figcompare}
\centering
\begin{minipage}{0.85\textwidth}
\includegraphics[width=\textwidth]{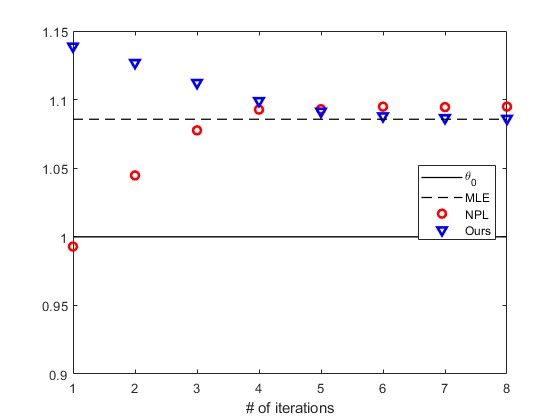}
\end{minipage}
\end{figure}

Table \ref{lambert} shows the simulation results of 1,000 replications with a sample size of 1,000. \SEES, MLE, and NPL perform very similarly. In particular, \SEES\ and MLE are almost identical in each replication. Figure \ref{figcompare} compares MLE and the iterations of NPL and \SEES\ in a typical replication.\footnote{\SEES\ usually converges in 2-4 iterations using our proposed choice of $\omega$. For better visualization in this figure, we increase $\log \omega$ in 7 equal steps to match the number of iterations of NPL.} While their earlier iterations could differ from MLE significantly, NPL and \SEES\ both converge to a close neighbourhood of MLE. 

In contrast, the two-step approach generates a larger bias and standard error. Alternatively, we can take the sample average in the second step. However, the noise in nonparametric estimates near boundaries deteriorates the estimates substantially. The median performs much better the mean. It is clear that the performance of the two-step estimator is affected by the first-stage nonparametric regression. 

\begin{table}[ht]
    \centering
    \caption{Monopoly Pricing: $\overline x=1$} \label{lambert}
\begin{tabular}{cccccc} \\ \hline \hline
     & \SEES   & MLE    & NPL    & \multicolumn{2}{c}{2-Step} \\ 
     &        &        &        & median       & mean        \\ \hline
mean & 1.0029 & 1.0030 & 1.0030 & 0.9639       & 1.4179      \\
se   & 0.1282 & 0.1283 & 0.1285 & 0.1356       & 2.8309     \\ \hline
\end{tabular}
\end{table}

\begin{rmk}
This simple model has a convenient property that each firm's optimal price is the solution of a strictly monotone function. In this case, the bisection method is fast and robust in solving the model, except that it takes many iterations to converge.\footnote{Alternatively, we show that the Lambert W function $W(x)$ can be calculated using the contraction mapping $\Psi(W,x) = x e^{-W}$ when $x$ is smaller than Euler's number.} In empirical games, the equilibrium is the solution of a system of nonlinear equations, which is harder to find and often lacks reliable algorithms. Appendix \ref{addsimu} provides additional simulations using a much richer DGP motivated by our empirical application of static games.
\end{rmk}

\section{Our Sieve-Based Efficient Estimator} \label{profile}


Now, we describe our estimator in detail. Instead of solving for $p^*(\theta)$ for each parameter $\theta$ in the likelihood evaluation, we approximate the true solution by $p^\beta$. The choice of approximation infrastructure depends on its approximation properties and computational convenience. A popular one often adopted in empirical studies is the method of sieves; that is, $p^\beta(\cdot)=\sum_{k = 1}^K \beta_{k}s_{k}(\cdot)$, where $\{s_1, \ldots, s_K\}$ represent the basis functions of the finite-dimensional sieve space $\mB$. Typical choices of $s_k$ include B-spline basis functions and Bernstein polynomials. Such methods are flexible in accounting for shape restrictions imposed by the structural model, such as nonnegativity and monotonicity. For instance, if $p$ represents choice probabilities, we can use $p^\beta(\cdot)=[1+\exp\{\sum_{k = 1}^K \beta_{k}s_{k}(\cdot)\}]^{-1}$ to ensure that $p\in (0,1)$.

Moreover, our \SEES\ imposes the model constraints by penalizing the difference between $p^\beta$ and $\Psi(p^\beta,\theta)$. This difference is independent of the data sample in measuring the fidelity of approximation to the equilibrium conditions. The smaller the difference, the better the approximate $p^\beta$ satisfies equilibrium conditions. 

We formulate the penalized log-likelihood criterion by combining the data fitting and model fitting criteria,
\begin{equation}
    \underbrace{\ell \big[p^\beta(\cdot), \theta \big]}_{\ell(\beta,\theta)} - \omega \times 
    \underbrace{\rho \Big[p^\beta(\cdot), \Psi \big(p^\beta(\cdot),\theta \big) \Big]}_{\rho(\beta,\theta)} , 
\label{pse-main}
\end{equation}
where $\omega >0$ is the smoothing parameter, and $\rho$ is a metric that measures the difference between $p^\beta$ and $\Psi(p^\beta,\theta)$. For instance, $\rho(p,\Psi(p,\theta)) = \|p - \Psi(p, \theta)\|_2^2$, where $\|\cdot\|_2$ is the Euclidean norm. For simplicity, we shall use as shorthand $\ell(\beta,\theta)$ and $\rho(\beta,\theta)$.

\subsection{Estimation Algorithms} \label{Sec:Algo}

We develop two algorithms to implement our estimator given each smoothing parameter $\omega$: a joint algorithm and a nested algorithm.

\vspace{.5cm}
\noindent
\textbf{Joint Algorithm} 
This algorithm is attractive because it involves a single-level optimization problem. 
We augment the unknown to $(\beta,\theta)$ and solve the following problem:
\begin{equation}
    \max_{\beta, \theta} \quad 
    \ell(\beta,\theta) - \omega \times 
    \rho(\beta,\theta) , 
\label{eq-joint}
\end{equation}
which leads to $\widehat{\beta}(\omega)$ and $\widehat{\theta}(\omega)$. We recommend supplying an analytic gradient and Hessian to reduce computational cost and to increase precision.

\vspace{.5cm}
\noindent
\textbf{Nested Algorithm}
This algorithm is intuitive, resembling MLE. There are two layers of optimization problems to be solved. In the inner layer, given $(\theta, \omega)$, we find the best approximation parameter $\widehat{\beta}(\theta; \omega)$ that solves the following problem: 
\begin{equation}
    \max_{\beta \in \Re^K} \quad 
    \ell(\beta,\theta) - \omega \times 
    \rho(\beta,\theta) .
\label{eq-inner}
\end{equation}
Solving \eqref{eq-inner} indicates that the maximizer $\widehat{\beta}(\theta; \omega)$ is an implicit function of $\theta$.

In the outer layer, applying the best fitting approximation parameter, we search for the structural parameter $\widehat{\theta}(\omega)$ that maximizes the following likelihood:
\begin{equation}
    \ell \left(\widehat{\beta}(\theta,\omega), \theta \right).
    \label{eq-outer}
\end{equation}
Note that we have considered the structural equation \eqref{eq-structure} in the inner layer. Therefore, the equilibrium conditions are embedded in $\widehat{\beta}(\theta, \omega)$. As illustrated above, the optimizer of \eqref{eq-inner}, $\widehat{\beta}(\theta; \omega)$, is an implicit function of $\theta$. Therefore, $\ell \left(\widehat{\beta}(\theta,\omega),\theta \right) $ in \eqref{eq-outer} is a function of $\theta$. We obtain the final estimator of $\theta$, denoted by $\widehat{\theta}(\omega)$, by directly maximizing \eqref{eq-outer} with respect to $\theta$.

\begin{rmk}
Because we solve the inner loop problem many times, it is more efficient to provide the gradient and Hessian of $h(\beta,\theta) = \ell(\beta,\theta) - \omega \rho(\beta, \theta)$ with respect to $\beta$, as well as the gradient of the objective function in the outer loop $\widehat{\ell}(\theta)=\ell(p^{\widehat{\beta}(\theta)}(\cdot),\theta)$ with respect to $\theta$. While the former is often straightforward, the latter is a bit more involved because the best-fitting approximation $\widehat{\beta}(\theta)$ is implicit. In particular, it requires deriving how the best-fitting approximation changes with respect to the model parameter, $\nabla \widehat{\beta}(\theta)$. Proposition \ref{prop-grad} provides the gradient of the outer loop.

Alternatively, we can use an \textit{alternating iterative} algorithm in place of either algorithm. That is, we can iterate the problem in \eqref{eq-inner} given a current estimate of $\theta$ and the problem in \eqref{eq-outer} to update the estimate of $\theta$. The iterations will continue until convergence. This iterative approach is similar to NPL.\footnote{Our algorithms, like other methods, do not guarantee finding global optima. We recommend experimenting with different starting points.}

\end{rmk}


\subsection*{The Choice of Tuning Parameters}

We propose a new method that selects the smoothing parameter $\omega$ based on the performance of parameter estimation. Intuitively, we choose a sufficiently large $\omega$ to ensure fidelity in approximating the equilibrium conditions.\footnote{Cross-validation is commonly used to balance bias and variance in estimation or prediction when the function of interest is unknown, by estimating prediction error or evaluating the likelihood function on held-out data. However, in our case, this trade-off is not a concern because the function $p(\cdot)$ is fully specified by the structural model \eqref{eq-structure}.} In particular, we start with a moderate $\omega_1$ and update it till the estimates converge. For each $\omega = \omega_\tau$, we can conduct the joint or nested algorithm and obtain an estimate $\widehat{\theta}(\omega_{\tau})$. Consider a significance level of $\alpha$. We obtain its standard error $\widehat{\sigma}(\omega_{\tau})$ and confidence interval $\mathcal{I}(\omega_{\tau}) = [\widehat{\theta}(\omega_{\tau}) +z_{\alpha/2} \widehat{\sigma}(\omega_{\tau}), \widehat{\theta}(\omega_{\tau}) +z_{1-\alpha/2} \widehat{\sigma}(\omega_{\tau})]$, applying the standard formula of standard error calculation for MLE. 

We multiply the smoothing parameter by $L$ each time, i.e. $\omega_{\tau+1} = L \times \omega_{\tau}$ and obtain a new estimate and its confidence interval $\mathcal{I}(\omega_{\tau+1})$. Continue this process till the overlapping portion of the two intervals accounts for more than a threshold percentage of both of the two. That is, our final choice of the smoothing parameter is $\widehat{\omega} = \omega_\tau$ if 
\[
\min \left\{
\frac{|\mathcal{I}(\omega_{\tau}) \cap \mathcal{I}(\omega_{\tau-1})|}{|\mathcal{I}(\omega_{\tau-1})|}, 
\frac{|\mathcal{I}(\omega_{\tau}) \cap \mathcal{I}(\omega_{\tau-1})|}{|\mathcal{I}(\omega_{\tau})|}
\right\} \geq \mathfrak{c} , 
\]
where $|\cdot|$ represents the length of the interval. In case the parameter of interest $\theta$ is multi-dimensional, we check this condition element by element. The final estimate of the model parameter follows $\widehat{\theta}(\widehat{\omega})$.

\begin{figure}[ht]
\caption{Choosing The Smoothing Parameter} \label{figlambda}
\centering
\begin{minipage}{0.85\textwidth}
\includegraphics[width=\textwidth]{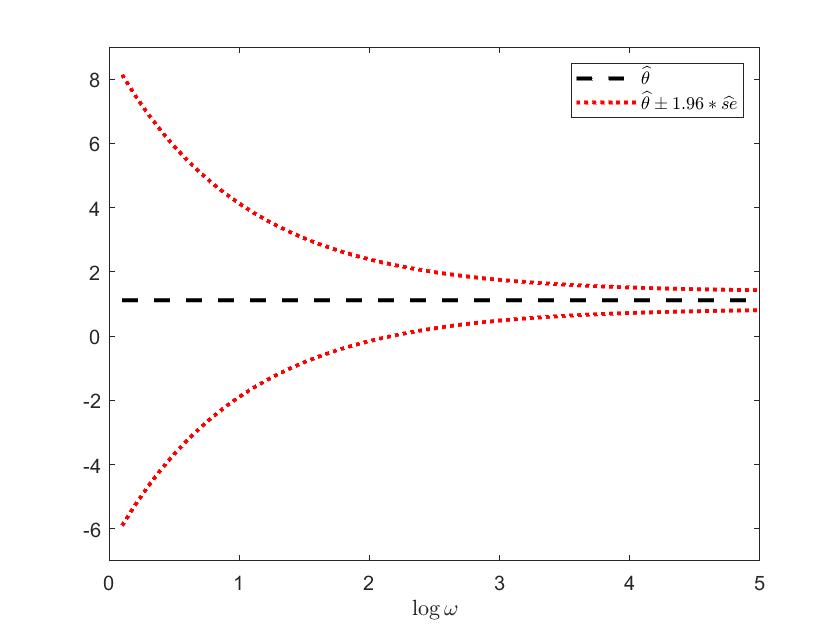}
{\footnotesize Note: The DGP is $y_j \sim \mathcal{N}(W(\theta_0),1)$, where $\theta_0 = 1$. This figure demonstrates how the estimate and its confidence interval change when the smoothing parameter increases.\par}
\end{minipage}
\end{figure}

All the simulations reported in this paper adopt $\alpha = 0.05$, $L = 10$, and $\mathfrak{c} = 95\%$. Therefore, $z_{\alpha/2} = -1.96$ and $z_{1-\alpha/2} = 1.96$. To illustrate how the proposed method works in terms of selecting the tuning parameter, consider the monopoly pricing example with $x_j = 1$. Figure \ref{figlambda} reports the parameter estimate and confidence intervals when the smoothing parameter varies. The x-axis represents $\log \omega$. While the bias seems small for small values of $\omega$, the confidence interval is large. As $\omega$ increases, it shrinks to the MLE confidence interval.

The approximation parameter $K$ (i.e., the number of basis functions) should be chosen properly: large enough to approximate the equilibrium well. Exactly how many is sufficient depends on the complexity of the equilibrium solution. In our motivating example, the solution is simple; as a result, we find that four cubic basis functions are adequate to approximate the solution well. When the solution is complex and the number of basis functions needs to be large, the analyst should start with a large smoothing parameter to avoid over-fitting in the inner loop; i.e., the likelihood function dominates. Sometimes, there are a finite number of states in the structural model, which is often assumed in estimating dynamic models. See, e.g., \citet{aguirregabiria2002swapping} and \citet{pesendorfer2008asymptotic}. Such finite states often come from discretization of covariates. In this case, the approximation can be perfect, i.e., $\beta = p$. That is, $s_k(x) = \mathbbm{1}(x = p_k)$, where $\mathbbm{1}(\cdot)$ is the indicator function and $p_k$ is the $k$th element in the endogenous variable $p$. Note in this scenario the number of basis functions is identical to the dimensionality of $p$. 

To capture empirically relevant covariates without losing much efficiency, any approximation methods would suffer from a computational curse of dimensionality --- the total number of basis functions has to grow fast as the dimensionality of $x$ increases. We propose to resolve this issue in several ways. First, more advanced approximation methods are often preferable to simple ones. See, e.g., \citet{chen2021efficient} compare neural networks-based estimators. Additional shape constraints, sparsity patterns, and better grid choices are useful in reducing computational burden. See, e.g., \citet{chen2007large} discusses various sieve-based methods, and  \citet{kristensen2021solving} discuss various approximation architectures for approximating value functions in dynamic models. Second, there are also many model-specific techniques for approximating functions using a small number of basis functions. The model may generate multiple endogenous objects, some directly observable while others intermediate. A well-chosen $p$ simplifies its approximation and evaluating $\Psi$ and data likelihood.  For instance, in static games of asymmetric information, if the deterministic component in the payoff function is linear in the parameters, see, e.g., \citet{bresnahan1991empirical} and \citet{bajari2010estimating}, how covariates determine the endogenous variable becomes a multiple-index model. In empirical auctions, \citet{chen2023identification} approximate the bid-stage primitives by flexible Bernstein polynomial sieves. One can borrow techniques from the existing literature in estimating such a model.

\subsection{Asymptotic Properties in Discrete State Settings} \label{asym}

As mentioned above, our proposed method can handle both continuous states, which result in an infinite-dimensional endogenous variable $p$, and discrete states, which lead to a finite-dimensional $p$. We begin by examining the discrete state settings, as this will provide insights into the continuous state cases.


Let $p = (p_1, \ldots, p_{d_1})^{\prime} \in \Re^{d_1}$ be the endogenous variable in \eqref{eq-structure}. Without loss of generality, we assume that $\theta \in \Re^{d}$ and $\Theta$ denotes the space of $\theta$. Under certain conditions that are specified below, we establish consistency and asymptotic normality of the joint estimator in the main text. The nested estimator is also consistent and asymptotically normal; see Theorems \ref{th-thetacons} and \ref{th-asynormal} in Appendix \ref{sec:nested}. In fact, the two estimators have the same asymptotic distribution. 

\begin{assump} \label{ass-compactness}
The parameter space $\Theta$ is a compact subset of $\Re^d$. 
\end{assump}

For any given $\theta \in \Theta$, we aim to maximize the following function with respect to $p$
\begin{equation}
    \ell_n(p) - \omega \|p - \Psi(p, \theta)\|_2^2,
\label{eq-obj}
\end{equation}
where $\ell_n$ denotes the log-likelihood corresponding to $n$ i.i.d. observations and $\|\cdot\|_2$ is the Euclidean norm of a vector. 
Suppose $Y_1, \ldots, Y_n$ are i.i.d. observations, taking values in $\Re^{d_2}$. 
We assume that the likelihood function in \eqref{eq-obj} can be written as
$$
\ell_n(p) = \frac{1}{n} \sum_{i = 1}^n f(Y_i, p),
$$
where $f$ is a function defined on $\Re^{d_2} \times \Re^{d_1}$.
For simplicity, we assume that $d_2 = 1$. In this context, $f(y, p)$ is actually the log density function of $Y_i$.

\begin{assump} \label{ass-fixed}
	There exists a compact and convex set
$K \subset \Re^{d_1}$ such that $p$ must lie in $K$.
\end{assump}

By Brouwer's fixed-point theorem, there must exist a solution to \eqref{eq-structure} for any $\theta \in \Theta$. For instance, $p\in [0,1]$ represents CCP in dynamic games. Define $g(p, \theta) = \Psi(p, \theta) - p$. Obviously, the solution to Equation \eqref{eq-structure}, denoted as $\ps(\theta)$, satisfies $g(\ps(\theta), \theta) = 0$. We impose the following regularity condition on $g(p, \theta)$.



\begin{assump} \label{ass-Lip}
    There exists a positive constant $r$ such that for any $(p_1, \theta), (p_2, \theta) \in K \times \Theta$ satisfying $\|g(p_1, \theta)\|_2 \vee \|g(p_2, \theta)\|_2  \leq r$, where $a \vee b$ denotes the larger value of $a$ and $b$, we have $\|p_1 - p_2 \|_2 \leq C\|g(p_1,\theta) - g(p_2, \theta)\|_2$ for some constant $C > 0$. 
\end{assump}


Assumption \ref{ass-Lip} can be understood as a local inverse Lipschitz condition. Consider the Lambert function $p = W(\theta)$, which is defined implicitly by $pe^{p} = \theta$. In correspondence, $g(p, \theta) =\theta e^{-p} - p$. Thus, $(p + g)e^{p + g} = \theta e^{g}$, which implies that $p + g = W(\theta e^{g})$ or
$$
p(g) = -g + W\left(\theta e^{g}\right). 
$$
Since $W$ is a continuously differentiable function by the implicit function theorem, for any $g_1, g_2$ satisfying $|g_1| \vee |g_2| \leq r$ with some constant $r$, we have $|p(g_1) - p(g_2)| \leq C|g_1 - g_2|$ for some constant $C$.


\begin{assump} \label{ass-ident}
	$\Psi$ is twice differentiable in both $p$ and $\theta$, and the Jacobian defined by $J_{g,p}(p, \theta) = \left[\frac{\p g_i}{\p p_j}(\ps(\theta), \theta)\right]$ is invertible. 
\end{assump}

By the implicit function theorem, this assumption ensures that the solution to Equation \eqref{eq-structure}, $p = \ps(\theta)$, is a continuously differentiable function of $\theta$. Let $\theta_0$ denote the true value of $\theta$. Define
$$
M(\theta) = \E_{\theta_0}[f(Y_i, \ps(\theta))] \quad\quad\text{for}~ \theta \in \Theta,
$$
where the expectation is taken with respect to $\Prob_{\theta_0}$.

\begin{assump} \label{ass-interior}
     The true value $\theta_0$ is in the interior of $\Theta$. 
\end{assump}

\begin{assump} \label{ass-unimax}
   $M(\theta)$ is a continuous function and has a unique maximum at $\theta_0$ in $\Theta$.  
\end{assump}
This assumption ensures that the true parameter $\theta_0$ is identifiable. 

\begin{assump} \label{ass-bounded}
	$f(y, p) \leq 0$ for any $(y, p) \in \Re^{1 + d_1}$ and $f(y, p) \in C(\Re \times \Re^{d_1})$.  Moreover, if $Y_i$ is not bounded, $f(y, p)$ satisfies that for any compact set $D \subset \Re^{d_1}$, 
	$$
	\liminf_{|y| \rightarrow \infty} \frac{1 + \inf_{x \in D} [-f(y, p)]}{\sup_{p \in D} [-f(y, p)]} > 0.
	$$
\end{assump}
This assumption holds for square loss functions, i.e., $f(y, p) = -(y - p)^2$. Given any $\theta \in \Theta$ and a positive $\omega$, recall the sieve estimate of $p$ is given by
\begin{equation} \label{eq-phat}
\hat{p}(\theta) = \argmax_{p} \frac{1}{n} \sum_{i = 1}^n f(Y_i, p) - \omega \|p - \Psi(p, \theta)\|_2^2. 
\end{equation}
The following theorem indicates the approximate solution to the structural equation \eqref{eq-structure} is uniformly close to the exact solution $\ps(\theta)$.


\begin{theorem} \label{th-pcons}
Assume that Assumptions \ref{ass-compactness}-\ref{ass-bounded} are satisfied. Then, we have
\begin{equation}
    \sup_{\theta \in \Theta} \|\hat{p}(\theta) - \ps(\theta) \|_2 = O_{\Prob}\left(\frac{1}{\sqrt{\omega_n}}\right),
    \label{eq-bound}
\end{equation}
provided that $\omega = \omega_n \rightarrow \infty$ as $n$ approaches infinity. 
\end{theorem}

\begin{rmk}
\label{rmk-bounded}

The error term originates from using a finite $\omega$. Since $\omega$ is finite, the solution to the penalized optimization problem in \eqref{eq-obj} is affected by the sample through the likelihood function $\ell_n$. Therefore, there exist discrepancies between this estimator and the solution to the structural equation \eqref{eq-structure}, which is the also the minimizer of the penalty term $\rho(p, \Psi(p, \theta))$ for any given $\theta$.


\end{rmk}

To establish the consistency of the estimator $\hat{\theta}_n$, we need a stronger version of Assumption \ref{ass-bounded}.

\begin{assump} \label{ass-strong}
	$f(y, p) \leq 0$ for any $(y, p) \in \Re^{1 + d_1}$ and $f(y, p) \in C(\Re \times \Re^{d_1})$ satisfies
	$$
	\E_{\theta_0} \left[\left\|\frac{\p f}{\p p}(Y_i, \ps(\theta_0))\right\|_2\right] < \infty. 
	$$  
	Moreover, if $Y_i$ is not bounded, $f(y, p)$ satisfies that for any compact set $D \subset \Re^{d_1}$, 
	$$
	\liminf_{|y| \rightarrow \infty} \frac{1 + \inf_{p \in D} [-f(y, p)]}{\sup_{p \in D} [-f(y, p)]} > 0~~\text{and}~~\liminf_{|y| \rightarrow \infty} \frac{1 + \inf_{p \in D} |\p f(y, p)/\p p_j|}{\sup_{p \in D} |\p f(y, p)/\p p_j|} > 0
	$$
	for $j = 1, \ldots, d_1$.
\end{assump}


Let $\tilde{\theta}_n$ denote the estimator of $\theta$ obtained from the joint algorithm. Actually, $\tilde{\theta}_n$ is defined by
\begin{equation} \label{eq-jointestimate}
  \tilde{\theta}_n = \argmax_{\theta \in \Theta} \ell_n(\hat{p}(\theta)) - w\|\hat{p}(\theta) - \Psi(\hat{p}(\theta), \theta)\|_2^2,
\end{equation}
where $\hat{p}(\theta)$ is given by \eqref{eq-phat}. 

\begin{theorem} \label{th-jointcons}
Suppose that Assumptions \ref{ass-compactness}-\ref{ass-unimax} and \ref{ass-strong} hold. 
If $\omega = \omega(n) \rightarrow \infty$, then $\tilde{\theta}_n$ is consistent. 
\end{theorem}

\begin{rmk} \label{rmk-consistency}
Since $\tilde{\theta}_n$ is the maximizer of $\ell_n(\hat{p}(\theta))$ with a non-positive penalty, we choose $\ell_n(\hat{p}(\theta))$ as the criterion function. Though it is difficult to evaluate the gradient of $\hat{p}(\theta)$ with respect to $\theta$, we can still use it as our criterion function, because we have established the bound on the difference between $\hat{p}(\theta)$ and $\ps(\theta)$ in Theorem \ref{th-pcons}. We establish the consistency and later the asymptotic normality of $\tilde{\theta}_n$ by resorting to the techniques for $M$-estimators. Even though $\tilde{\theta}_n$ is not the maximizer of $\ell_n(\ps(\theta))$, we are able to control the difference between $\ell_n(\ps(\tilde{\theta}_n))$ and $\max_{\theta \in \Theta}\ell_n(\ps({\theta}))$. We show that, in actuality, $\tilde{\theta}_n$ nearly maximizes $\ell_n(\ps({\theta}))$, and then we leverage the results for $M$-estimators; see Section 5.2 of \cite{van1996} for more details.\end{rmk}

To derive asymptotic normality, we need a stronger condition than Assumptions \ref{ass-bounded} and \ref{ass-strong}. 

\begin{assump} \label{ass-normal}
	$f(y, p) \leq 0$ for any $(y, p) \in \Re^{1 + d_1}$ and $f(y, p) \in C(\Re \times \Re^{d_1})$ satisfies
	$$
	\E_{\theta_0} \left[\left\|\frac{\p f}{\p p}(Y, \ps(\theta_0))\right\|_2^2\right] < \infty \quad\text{and}\quad 
	\E_{\theta_0} \left[\left|\frac{\p^2 f}{\p p_i \p p_j}(Y, \ps(\theta_0))\right|\right] < \infty
	$$
	for $i, j = 1, \ldots, d_1$. 
	Moreover, if $Y_i$ is not bounded, $f(y, p)$ satisfies that for any compact set $D \subset \Re^{d_1}$, 
	\begin{align*}
	\liminf_{|y| \rightarrow \infty} \frac{1 + \inf_{p \in D} [-f(y, p)]}{\sup_{p \in D} [-f(y, p)]} > 0~~~~\liminf_{|y| \rightarrow \infty} \frac{1 + \inf_{p \in D} |\p f(y, p)/\p p_j|}{\sup_{p \in D} |\p f(y, p)/\p p_j|} > 0 
	\end{align*}
	for $j = 1, \ldots, d_1$, and
	$$
	\liminf_{|y| \rightarrow \infty} \frac{1 + \inf_{p \in D} |\p^2 f(y, p)/\p p_i \p p_j|}{\sup_{p \in D} |\p^2 f(y, p)/\p p_i\p p_j|} > 0 
	$$
	for any $i, j = 1, \ldots, d_1$. 
\end{assump}

\begin{theorem} \label{th-jointasynormal}
Suppose that Assumptions \ref{ass-compactness}-\ref{ass-unimax} and \ref{ass-normal} hold. 
If $\omega_n/n^2 \rightarrow \infty$ and the matrix
$$
V_{\theta_0} = \E_{\theta_0} \left[\left\{\frac{\p \ps(\theta_0)}{\p \theta}\right\}^{\prime} \frac{\p^2 f(Y_i, \ps(\theta_0))}{\p p \p p^{\prime}} \left\{\frac{\p \ps(\theta_0)}{\p \theta}\right\}\right]
$$
is invertible, then 
$$
\rt{n}(\tilde{\theta}_n - \theta_0) \overset{d}{\to}
\mN(0, \Sigma),
$$
where $\Sigma = V_{\theta_0}^{-1} \left(\frac{\p  \ps(\theta_0)}{\p \theta}\right)^{\prime} \E_{\theta_0} \left[\left\{\frac{\p f(y, \ps(\theta_0))}{\p p}\right\}\left\{\frac{\p f(y, \ps(\theta_0))}{\p p}\right\}^{\prime}\right]\left(\frac{\p  \ps(\theta_0)}{\p \theta}\right) V_{\theta_0}^{-1}.$
\end{theorem}

\begin{rmk}  \label{rmk-mle}
Under mild regularity conditions on $f$, 
$$
- \E_{\theta_0} \left[\frac{\p^2 f(Y_i, \ps(\theta_0))}{\p p \p p^{\prime}}\right] = \E_{\theta_0} \left[\left\{\frac{\p f(y, \ps(\theta_0))}{\p p}\right\}\left\{\frac{\p f(y, \ps(\theta_0))}{\p p}\right\}^{\prime}\right],
$$
then $\Sigma = -V_{\theta_0}^{-1}$. Consequently, 
even though $\tilde{\theta}_n$ is different from the maximum likelihood estimator of $\theta$, which minimizes $\ell_n(\ps(\theta))$, the asymptotic variance of $\hat{\theta}_n$ attains the Cram\'er-Rao lower bound. Thus, $\tilde{\theta}_n$ is asymptotically efficient. 
\end{rmk}

The joint algorithm is attractive because it involves a single-level optimization problem and computes the Hessian matrix with respect to $(\beta,\theta)$ at the solution directly. The following corollary provides a natural way to calculate the standard error of $\tilde{\theta}_n$ using the Hessian matrix generated from the joint algorithm.

\begin{corollary} \label{fisherjoint}
The Fisher information can be characterized as 
\[
\widehat{H} = \bf{H}_{\theta \theta} - \bf{H}_{\beta \theta}^\prime \bf{H}_{\beta \beta}^{-1} \bf{H}_{\beta \theta}, 
\]
where the matrices in bold are the four blocks in the Hessian matrix generated from a joint maximization algorithm,
\begin{align*}
        \begin{bmatrix}
\bf{H}_{\beta \beta}  & \bf{H}_{\beta \theta} \\
\bf{H}_{\theta \beta}  & \bf{H}_{\theta \theta} 
\end{bmatrix} .
\end{align*}
\end{corollary}



\subsection{Asymptotic Properties in Continuous State Settings} \label{asym-semi}

In this section, we consider continuous states in the structural model and examine the large-sample properties of SEES.

Let $x$ denote the continuous states. Without loss of generality, we assume that the dimension of $x$ is 1. 
Given any $\theta \in \Theta \subset \Re^d$, $p^*(x; \theta) \in \Re^{d_1}$ denotes the solution to the following the structure equation:
\begin{equation}
    p(x) = \Psi(p(x), \theta),
    \label{eq-structure-semi}
\end{equation}
where $x \in [0, T]$ with a fixed $T > 0$. 
Without loss of generality, we assume $T = 1$. 
We approximate the solution to \eqref{eq-structure-semi} using a sieve method. In particular, we take the sieve space, denoted as $\mB_n$, to be the space of cubic B-spline functions equipped with knots $\tau^{(n)} = \left\{0 = t_1^{(n)} < \cdots < t_{M_n}^{(n)} = T\right\}$. Let $|\tau^{(n)}| = \max_{1 \leq i \leq M_n - 1} |t_{i + 1}^{(n)} - t_{i}^{(n)}|$ be the largest distance of adjacent knots in $\tau^{(n)}$. For any element $\eta \in \mB_n$, there exists a $\beta = (\beta_1, \ldots, \beta_{K_n})^{\top} \in \Re^{K_n}$ such that $\eta(x) = \sum_{j = 1}^{K_n} \beta_j s_j(x)$, where $s_j$'s are cubic B-spline basis functions and $K_n = M_n + 3$.

Suppose that $(Y_1, X_1), \ldots, (Y_n, X_n)$ are i.i.d observations, where $X_i$'s are independently sampled from a distribution $Q$ on $[0, T]$ and $Y_i$'s take values in $\Re^{d_2}$. For simplicity, we assume that $d_1 = d_2 = 1$. 
Following the notations defined in the last subsection, given $\theta \in \Theta$, we have
\begin{equation}
   \hat{p}_n(\cdot; \theta) = \argmax_{p \in \mB_n} \ell_n(p(\cdot)) - \omega \rho[p(\cdot), \Psi(p(\cdot), \theta)],
\label{eq-obj-semi}
\end{equation}
where the likelihood $\ell_n$ can be written
$$
\ell_n(p(\cdot)) = \frac{1}{n} \sum_{i = 1}^n f(Y_i, p(X_i)),
$$
for any function $p$ defined on $[0, T]$, and the penalty function $\rho$ is given by 
$$
\rho[p(\cdot), \Psi(p(\cdot), \theta)] = \int_0^T \{p(x) - \Psi(p(x), \theta)\}^2 \diff x.
$$
Then the nested estimator of $\theta$ is given by
\begin{equation}
\hat{\theta}_n = \argmax_{\theta \in \Theta} \ell_n(\hat{p}_n(\cdot; \theta)).
\end{equation}
We next study the asymptotic properties of $\hat{\theta}_n$ in this model. Similar to Section \ref{asym}, we first establish consistency for the estimator and then develop the asymptotic normality of $\hat{\theta}_n$. To this end, we need the following regularity conditions.

\begin{assumpB} \label{ass-density}
	The density function of $Q$, denoted by $q$, satisfies that $C_1 \leq q(x) \leq C_2$ for any $x \in [0, T]$, where $C_1$ and $C_2$ are two positive constants. 
\end{assumpB}

Assumption \ref{ass-density} ensures that $X_i$'s are evenly distributed over $[0, T]$, which is entailed by a good estimation of $p$ over the entire domain. Moreover, this assumption is commonly adopted in the literature of nonparametric smoothing; see \cite{stone1985additive} and \cite{chen2023identification} for example. 

\begin{assumpB} \label{ass-compactness-semi}
The parameter space $\Theta$ is a compact subset of $\Re^d$. 
\end{assumpB}

Define $g(p(x), \theta) = \Psi(p(x), \theta) - p(t)$ for any function $p$ defined on $[0, T]$. Let $h^{(k)}$ denote the $k$th order derivative of function $h$ for any integer $k \geq 0$, and define
$$
C^k([0, T]) = \{h: h^{(k)}~\text{is continuous on}~[0, T] \}.
$$
The following two assumptions extend their counterparts in the parametric model to the semi-parametric one. 
\begin{assumpB} \label{ass-ident-semi}
For any $\theta \in \Theta$, $\Psi \in C^4(\Re \times \Theta)$. 
	  There exists a positive constant $r$ such that for any $(p_1, \theta), (p_2, \theta)$ satisfying $\max\{\|p_1\|_{\infty}, \|p_2\|_{\infty}\} \leq r$ and $\theta \in \Theta$, we have $\|p_1 - p_2 \|_{L^2([0, T])} \leq C_g\|g(p_1,\theta) - g(p_2, \theta)\|_{L^2([0, T])}$ for some constant $C_g > 0$. 
\end{assumpB}



By Lemma \ref{le-appro-semi} in Appendix \ref{app-semi}, we define the sieve space to be 
$$
\mB_n(r) = \left\{\eta(x): \eta(x) = \sum_{j = 1}^{K_n} \beta_j s_j(x), \|\eta\|_{\infty} \leq r \right\}. 
$$
with equally spaced knots for some sufficiently large constant $r$. 
Therefore, for any $\theta \in \Theta$, there exits an $p_{\theta, n} \in \mB_n(r)$ such that $\|p^*(\cdot; \theta) - p_{\theta, n}\|_{\infty} = O(K_n^{-4})$. Let 
\begin{equation}
    r_n = \sup_{\theta \in \Theta} \inf_{\eta \in \mB_n(r)}  \|p^*(\cdot; \theta) - \eta\|_{\infty}. 
    \label{eq-rn}
\end{equation}
Similar to Section \ref{asym}, for the sieve estimator $\hat{p}(\cdot; \theta)$ defined in \eqref{eq-obj-semi}, we establish an important approximation error bound, which will be used to develop the asymptotic normality for $\hat{\theta}_n$ later. 

\begin{theorem} \label{th-pcons-semi}
Assume that Assumptions \ref{ass-density}-\ref{ass-ident-semi} and \ref{ass-strong-semi} are satisfied. Furthermore, if $\omega = \omega_n \rightarrow \infty$ as $n \rightarrow \infty$,  we have
\begin{equation*}
  \sup_{\theta \in \Theta}  \|\hat{p}_n(\cdot; \theta) - p^*(\cdot; \theta)\|_{L^2(\Prob)}^2 : = \sup_{\theta \in \Theta} \int_0^T  \{\hat{p}_n(x; \theta) - p^*(x; \theta)\}^2 q(x) \diff x = O_{\Prob}(\omega_n^{-1}) +  O_{\Prob}(r_n^2). 
\end{equation*}
\end{theorem}

Let $\theta_0$ denote the true value of $\theta$ and $G_{\theta_0}$ denote the joint distribution of $(X_i, Y_i)$ under this true value. Define
$$
M(\theta) = \E_{\theta_0}[f(Y_i, p^*(X_i; \theta)], \quad\quad \theta \in \Theta,
$$
where the expectation is taken with respect to $G_{\theta_0}$. The follow assumptions are essentially the same as those in Section \ref{asym}.

\begin{assumpB} \label{ass-unimax-semi}
	$M(\theta)$ is continuous function and has a unique maximum at $\theta_0$ in $\Theta$.  
\end{assumpB}

\begin{assumpB} \label{ass-strong-semi}
	$f(y, p) \leq 0$ for any $(y, p) \in \Re^2$ and $f(y, p) \in C(\Re \times \Re)$ satisfies
	$$
	\E_{\theta_0} \left[\left|\frac{\p f}{\p p}(Y_i, \ps(X; \theta_0))\right|\right] < \infty. 
	$$  
	Moreover, if $Y_i$ is not bounded, $f(y, p)$ satisfies that for any compact set $D \subset \Re$, 
	$$
	\liminf_{|y| \rightarrow \infty} \frac{1 + \inf_{p \in D} [-f(y, p)]}{\sup_{p \in D} [-f(y, p)]} > 0~~\text{and}~~\liminf_{|y| \rightarrow \infty} \frac{1 + \inf_{p \in D} [-\p f(y, p)/\p p]}{\sup_{p \in D} [-\p f(y, p)/\p p]} > 0
	$$
\end{assumpB}

To establish consistency and asymptotic normality for $\hat{\theta}_n$, a stronger version of Assumption \ref{ass-strong-semi} is entailed.

\begin{assumpB} \label{ass-normal-semi}
	$f(y, p) \leq 0$ for any $(y, p) \in \Re^{2}$ and $f(y, p) \in C(\Re \times \Re)$ satisfies
 that under $G_{\theta_0}$, ${\p f}(Y_i, \ps(X_i; \theta_0))/{\p p}$ is sub-Gaussian, and
	$$
	\quad 
	\E_{\theta_0} \left[\left|\frac{\p^2 f}{\p p^2}(Y, \ps(X; \theta_0))\right|\right] < \infty.
	$$
Moreover, if $Y_i$ is not bounded, $f(y, p)$ satisfies that for any compact set $D \subset \Re$, 
	\begin{align*}
	\liminf_{|y| \rightarrow \infty} \frac{1 + \inf_{p \in D} [-f(y, p)]}{\sup_{p \in D} [-f(y, p)]} > 0~~~~\liminf_{|y| \rightarrow \infty} \frac{1 + \inf_{p \in D} |\p f(y, p)/\p p|}{\sup_{p \in D} |\p f(y, p)/\p p|} > 0,
	\end{align*}
 and
	$$
	\liminf_{|y| \rightarrow \infty} \frac{1 + \inf_{p \in D} |\p^2 f(y, p)/\p p^2 |}{\sup_{p \in D} |\p^2 f(y, p)/\p p^2|} > 0.
	$$
	
\end{assumpB}

\begin{rmk}
\label{rmk-subgaussian-semi}
 We impose a sub-Gaussian condition on ${\p f}(Y_i, \ps(X_i; \theta_0))/{\p p}$ in Assumption \ref{ass-normal-semi}, which is stronger than that in Assumption \ref{ass-normal}. This term is usually referred to as the residual in the gradient descent algorithm. Actually, this condition is met when $\epsilon_i = Y_i - p(X_i)$ follows a normal distribution and $f(Y_i, p(X_i)) = -(Y_i - p(X_i))^2$ or in logistic regression when $\Prob(Y_i = 1 | X_i) = \exp\{p(X_i)\}/[1 + \exp\{p(X_i)\}]$. We impose this condition to ensure that $\ell_n(\ps(\cdot; \hat{\theta}_n)) \geq \sup_{\theta \in \Theta} \ell_n(\ps(\theta)) - o_{\Prob}(n^{-1})$. Alternatively, we may impose a stronger smoothness condition on $\ps(\cdot; \theta)$. Then a smaller approximation error, i.e., a smaller $r_n$ (defined in \eqref{eq-rn}), can be obtained with a sieve space with higher-order B-spline functions. This reflects a trade-off between the smoothness of the structure equation and the decaying rate of the tail probability of ${\p f}(Y_i, \ps(X_i; \theta_0))/{\p p}$.
\end{rmk}

\begin{theorem} \label{th-consistency-semi}
Suppose Assumptions \ref{ass-density}-\ref{ass-unimax-semi} and \ref{ass-normal-semi} hold. 
If $\omega_n \rightarrow \infty$ and $K_n^2 \log (K_n) = o(n)$, then $\hat{\theta}_n$ is a consistent estimator of $\theta_0$. 
\end{theorem}

\begin{theorem} \label{th-nestasynormal-semi}
Suppose that Assumptions \ref{ass-density}-\ref{ass-unimax-semi} and Assumption \ref{ass-normal-semi} hold. 
If $\omega_n/n^2 \rightarrow \infty, n^{1/4} = o(K_n)$, $K_n^2 \log(K_n) = o(n)$ and the matrix
\begin{align*}
\begin{split}
& V_{\theta_0} = -\E_{\theta_0} \left[\frac{\p f(Y_i, \ps(X_i; \theta_0))}{\p p} \frac{\p^2 \ps(X_i; \theta_0)}{\p \theta \p\theta^{\prime}} \right. \\
& \quad \left. +  \frac{\p^2 f(Y_i, \ps(X_i; \theta_0))}{\p p^2} \left\{\frac{\p \ps(X_i; \theta_0)}{\p \theta}\right\} \left\{\frac{\p \ps(X_i; \theta_0)}{\p \theta}\right\}^{\prime}\right]
\end{split}
\end{align*}
is invertible, then 
$$
\rt{n}(\hat{\theta}_n - \theta_0) \overset{d}{\to}
\mN(0, \Sigma),
$$
where $\Sigma = V_{\theta_0}^{-1}  \E_{\theta_0} \left[\left\{\frac{\p f(Y, \ps(X; \theta_0))}{\p p}\right\}^2 \left\{\frac{\p \ps(X_i; \theta_0)}{\p \theta}\right\}^{\prime} \left\{\frac{\p \ps(X_i; \theta_0)}{\p \theta}\right\}\right]V_{\theta_0}^{-1}.$
\end{theorem}

\begin{rmk}
\label{rmk-normality-semi}
1) If the sieve space $\mB_n(r)$ consists of the cubic spline functions equipped with equally spaced knots $\tau^{(n)}$, then by Lemma \ref{le-appro-semi}, this condition implies $r_n = o(n^{-1})$. Meanwhile, we impose an upper bound on $K_n$ to control the bracketing number \citep[cf.][p. 270]{van2000} of a relevant functional class when we study the uniform estimation error of $\hat{p}_n(\cdot, \theta)$ relative to $\ps(\cdot, \theta)$ over $\theta \in \Theta$. 
2)Using the same technique as in Section \ref{asym}, we can easily show that the joint estimator $\tilde{\theta}_n$, which is defined as in \eqref{eq-jointestimate} in this setting, has the same limiting distribution under the conditions of Theorem \ref{th-nestasynormal-semi}. 3) We approximate the solution by flexible sieves so that the approximation error disappears in first order asymptotics.
\end{rmk}

\subsection{Discussion}


\noindent \textbf{MPEC: } 
When the state space is discrete and $p$ is finite, our method could incorporate each element of the endogenous variable $p$ as a basis function in sieve approximation and put all of the weight on the equilibrium conditions in the inner loop. In this case, our estimator becomes the MPEC estimator. 

Moreover, it is easy to show that for a given $\omega >0$, there exists an $\epsilon>0$ such that the optimization problem of the joint algorithm \eqref{eq-joint} has the same solution as its dual optimization problem
\begin{align*}
    \max_{\beta, \theta} & \quad 
    \ell(\beta,\theta) \\
    \textit{s.t.} & \quad \rho(\beta,\theta) \leq \epsilon
\end{align*}
The dual problem is a natural generalization of the MPEC estimator. However, solving it numerically is challenging. 


Our above-mentioned derivation also suggests a natural way to calculate standard errors for MPEC estimators. 


\begin{corollary}\label{MPEChessian}
When $\omega = \infty$ and $\beta = p$, the observed Fisher information can be characterized as  
\[
\bf{\widehat{H}} = \bf{H}_{\theta \theta} 
+ [\nabla \widehat{\beta}(\widehat{\theta})]^\prime 
\bf{H}_{\beta \beta}[\nabla \widehat{\beta}(\widehat{\theta})] +
[\nabla \widehat{\beta}(\widehat{\theta})]^\prime \bf{H}_{\beta \theta}  + 
[\bf{H}_{\beta \theta}]^\prime \nabla \widehat{\beta}(\widehat{\theta}) , 
\]
where $\nabla\beta(\widehat{\theta})=-\left[\nabla_\beta g\right]^{-1} \times \nabla_\theta g$ on the right-hand side, and the matrices in bold are the four blocks in the Hessian matrix generated from a constrained maximization algorithm,
\begin{align*}
        \begin{bmatrix}
\bf{H}_{\beta \beta}  & \bf{H}_{\beta \theta} \\
\bf{H}_{\theta \beta}  & \bf{H}_{\theta \theta} 
\end{bmatrix} .
\end{align*}
\end{corollary}

To the best of our knowledge, this result is new in the literature. \citet{su2012constrained} suggest obtaining standard errors through bootstrapping. We derive the general result in Section \ref{asym}. Here, we consider $p,\beta,\theta \in \Re$, as in the simple example with $x_j=1$, to explain the idea. When $\omega = \infty$ and $\beta = p$, our estimator is effectively an MPEC estimator,
\[
\max_{g(\beta,\theta) =0} \quad \ell(\beta, \theta).
\]
The MPEC approach forms the Lagrangian function $h(\beta,\theta,\omega) = \ell (\beta, \theta ) + \lambda g(\beta, \theta)$. Note that this multiplier $\lambda$ should not to be confused with the smoothing parameter $\omega$ for general PSE. By definition, we have $g(\widehat{\beta}(\theta),\theta ) = 0$. Its first-order and second-order derivatives are
\begin{align*}
    g\lo\beta \widehat{\beta}^\prime(\theta) + g\lo\theta & = 0 \\
    g\lo{\beta \beta} [\widehat{\beta}^\prime(\theta)]^2 + 2 g\lo{\beta \theta} \widehat{\beta}^\prime(\theta) + g\lo\beta \widehat{\beta}^{\prime\prime}(\theta) + g\lo{\theta \theta} & = 0,
\end{align*} 
which allow for expressing $\widehat{\beta}^{\prime}(\theta)$ and $\widehat{\beta}^{\prime\prime}(\theta)$ in the gradient of $g$. 

On the other hand, the second-order derivative of the likelihood is
\begin{align*}
        \widehat{\ell}_{\theta\theta} (\theta) |_{\theta = \widehat{\theta}} & =
    \ell_{\beta \beta} [\widehat{\beta}^\prime(\theta)]^2 + 2 \ell_{\beta \theta} \widehat{\beta}^\prime(\theta) + \ell_\beta \widehat{\beta}^{\prime\prime}(\theta) + \ell_{\theta \theta} \\
    & = \Big[ \left(\frac{g\lo\theta}{g\lo\beta}\right)^2  (\ell_{\beta \beta} + \lambda g\lo{\beta \beta})  
    -2 \frac{g\lo\theta}{g\lo\beta} (\ell_{\beta \theta} + \lambda g\lo{\beta \theta})
    + (\ell_{\theta \theta} + \lambda g\lo{\theta \theta})
    \Big]\bigg|_{\beta = \widehat{\beta}(\widehat{\theta}) , \theta = \widehat{\theta}},
    \end{align*}
    where $\lambda$ denotes the associated Lagrange multiplier reported by a constrained maximization algorithm. The last equation follows from the Lagrange multiplier theorem that $\ell_\beta + \lambda g\lo\beta =0$ at the optimum $(\beta = \widehat{\beta}(\widehat{\theta}) , \theta = \widehat{\theta})$ and the first-order and second-order derivatives of the equilibrium constraints. All terms on the RHS are readily available if MPEC converges. We recommend supplying the analytic gradient and Hessian, as the numerical one can be inaccurate.


To the best of our knowledge, the theoretical properties of the MPEC estimator for structural models with continuous states remain unexplored. In contrast, our proposed method offers a rigorous framework for conducting statistical inference for $\theta$ with either discrete, continuous, or both types of states. We believe this represents a critical advancement for practical applications.


\vspace{1cm}
\noindent \textbf{Approximate MLE: } 
We now discuss the extreme case when we let $\omega = \infty$. That is, for each guess of the model parameter $\theta$, we find the best approximation to minimize any deviation from the equilibrium condition and then evaluate the likelihood by plugging in this best approximation. Specifically, our estimator becomes equivalent to  
\begin{align*}
    \max_{\theta} & \quad 
\ell(p^{\beta(\theta)},\theta; \text{data}) \\
\text{where } & \quad  \beta(\theta) = \arg \min_{\beta} \rho(p^\beta,\Psi(p^\beta,\theta)),
\end{align*}
which looks similar to MLE, with an important difference that we only search for the best approximation in the inner loop. We call this special case of our estimator the approximate MLE (AMLE). Such approximate solution approaches have appeared in the literature. See, e.g., \citet{keane1994solution,keane1997career} use sieves to approximate solutions in dynamic structural models.


One may wonder about the advantages of gradually changing $\omega$ instead of directly considering the limiting case. AMLE ignores the data when finding the best approximation of the solution for each $\theta$. Because the data are informative about the true strategies $p$, our general sieve-based efficient estimator may perform better than AMLE. By gradually updating the smoothing parameter, we shift the weight from the data to the equilibrium condition. At the minimum, a preliminary nonparametric estimate of $\widehat{p}$ (by letting $\omega =0$) constitutes a good starting value for the inner loop but is subject to issues with nonparametric estimates. When the smoothing parameter increases, more weight is given to the equilibrium condition. By forcing model restrictions more strongly, the estimates converge to MLE estimates. 


\section{Application: Walmart versus Kmart Entry Game} \label{sec:app}

In this section, we apply our methodology to an entry game between Walmart and Kmart, using a dataset published by \citet{jia2008happens}. A detailed description of the industry and data is available in the original paper.

\subsection{Data}

The original dataset includes 2,065 markets, each representing a county with an average population ranging from 5,000 to 64,000, covering the years 1988 to 1997. For our analysis, we focus on the year 1997. The market-level variables include the log of county population (pop), the log of retail sales per capita (spc), and the percentage of urban population (urban). Walmart-specific variables include an intercept, the log of distance to Bentonville (dbenton), and an indicator for the southern region. Kmart-specific variables include an intercept and an indicator for the Midwest region (midwest). These variables capture key variations in the data. For instance, a simple scatter plot of the total number of firms shows that neither firm enters the market when SPC is too low. 


Denote the data as $\{d_{\W m}, d_{\K m}, x_{\W m}, x_{\K m}, z_m \}_{m=1}^M$, where $\W$ and $\K$ represent Walmart and Kmart, respectively. Here, $d_{jm}$ is firm $j$'s entry decision in market $m$, $x_{jm}$ includes firm-specific covariates, including a constant, and $z_m$ contains market-specific covariates. Table \ref{SumJia2008} provides summary statistics for the sample used in our analysis.

\begin{table}[ht] \label{SumJia2008}
\centering
\begin{threeparttable}
\caption{Summary Statistics}
\begin{tabular}{lrrrr} \\ \hline 
Variable   & Mean & Std. Dev. & Min  & Max   \\  \hline 
$d_{\W m}$    & 0.48 & 0.50      & 0    & 1     \\
$d_{\K m}$      & 0.19 & 0.39      & 0    & 1     \\  \hline 
pop & 2.98 & 0.67      & 1.54 & 4.37  \\
spc        & 8.20 & 0.47      & 5.08 & 10.66 \\
urban      & 0.33 & 0.24      & 0    & 1     \\
dbenton    & 6.24 & 0.63      & 3.01 & 8.29  \\  
southern   & 0.50 & 0.50      & 0    & 1     \\ 
midwest    & 0.42 & 0.49      & 0    & 1    \\ \hline  
\end{tabular}
\end{threeparttable}
\end{table}

\subsection{Empirical Model}

For the purpose of illustrating our method, we model the entry game between Walmart and Kmart as a static game with incomplete information. Two players, Walmart ($\W$) and Kmart ($\K$), decide whether to enter a market. We assume that they make independent decisions across markets. Let $d_{j}=1$ if firm $j$ is active and $0$ otherwise. The payoff function of firm $j$ depends on its own productivity, whether its competitor enters or not, market- and firm-specific covariates, and private information:
\[
u_{j}(d_{j},d_{-j})=\underbrace{X_{j}^\prime \beta - Z^\prime \gamma}_{\xi_{j}} - \Delta d_{-j} + \epsilon_{j1} ,
\]
if $d_{j}=1$ and $=\epsilon_{j0}$ otherwise, where $X =(X_{\mathcal{W}},X_{\mathcal{K}})^{\prime}$ is firm characteristics that affect only the focal firm's profit and $Z$ is market characteristics common to both firms. For convenience, we denote $\xi_{j} = X_{j}^\prime \beta - Z^\prime \gamma$. 

Firm $j$'s profit is $\xi_j$ under monopoly and $\xi_j - \Delta$ under duopoly. Note we allow asymmetry in monopoly profit by including a constant in firm-specific covariates. The term $Z^\prime \gamma$ is common among all firms. Denote $\theta = (\beta, \gamma, \Delta)^\prime$, market- and firm-specific characteristics $(x,z)$ are common knowledge, and firm $j$'s private information $\epsilon_{j}$ is type-1 extreme value distributed and independent of $\epsilon_{-j}$.

Therefore, the probability that firm $j$ chooses to enter is 
\[
p_{j} = \frac{1}{1+\exp \{- \xi_{j} + p_{-j} \Delta\}} ,
\]
where $p_{-j}$ is its competitor's entry probability. Denote the CCPs as $p=(p_{\W},p_{\K})^{\prime}$. Define the best response mapping from CCP to CCP $\Psi: p\rightarrow p$. In equilibrium, we must have 
\[
p=\Psi(p,\theta).
\]

We define the likelihood function as 
\[
\ell(p^\beta, \theta)= 
\sum_{j=\W,\K}\sum_{m=1}^{M}\left\{ 
d_{jm} \log \Big[p_{j}^\beta(\chi_{m}) \Big] 
+(1-d_{jm} )\log \Big[1-p_{j}^\beta(\chi_{m}) \Big] \right\},
\]
where $\chi_m = (x_{\W m}, x_{\K m}, z_m)^\prime$, the approximation structure $p_j^\beta$ will be introduced below, and the penalization as 
\[
\rho(\beta, \theta) = \sum_{j=\W,\K}\sum_{m=1}^{M} \left[p_j^\beta(\chi_m) - \Psi_j\left(p_{-j}^\beta(\chi_m) , \theta\right)\right]^2 ,
\]
which accounts for the equilibrium conditions for the set of observed market-specific covariates. In addition, we supply analytic gradient; see Appendix \ref{JHjia}.


\vspace{1cm}
\noindent \textbf{Approximation Structure: } We now consider the approximation of the CCPs. A naive approach is to approximate them as a flexible function of all market- and firm-specific covariates $p_j(x_j, x_{-j},z)$, which is of six dimensions in our empirical setting. To ensure that the approximation error disappears in first order asymptotics, the dimension of the approximation parameter $K$ needs to be large, leading to substantial computational challenges.

We propose a novel approximation structure that leverages the model structure: the deterministic component in the payoff function is linear in the parameters. As a result, how covariates $(x,z)$ determine the endogenous variable $p$ becomes a two-index model $p^*(\xi_j, \xi_{-j})$, which is much easier to approximate than a six-dimensional function. Using cubic basis functions following \citet{luo2018structural}, we propose to approximate the CCPs in our empirical model by 
\begin{equation} \label{oursieve}
    p^\beta (\xi_{j},\xi_{-j}) = \sigma \Bigg(\sum_{\imath=1}^K\sum_{\jmath=1}^K\beta_{\imath \jmath} s_\imath \big(\sigma(\xi_{j}) \big) s_{\jmath} \big(\sigma(\xi_{-j}) \big) \Bigg), 
\end{equation}
where $s_\imath(\cdot)$ and $s_\jmath(\cdot)$ are cubic spline basis functions on $[0,1]$, $\beta = (\beta_{11},\ldots, \beta_{K K})^\prime$ and $\sigma(\cdot)  = (1+e^{-\cdot})^{-1}$ representing the logistic function. In principle, we can use a different number of basis functions in the two dimensions. For convenience, we will use the same number $K$ and refer to it as the approximation parameter.

Note that the logistic function appears three times but for different reasons. First, because $s_\imath(\cdot)$ and $s_\jmath(\cdot)$ are cubic spline basis functions on $[0,1]$, the inner ones $\sigma(\xi_j)$ and $\sigma(\xi_{-j})$ transforms unbounded payoff indices $\xi_j$ and $\xi_{-j}$ into bounded ones on $[0,1]$. Interestingly, there are just the stand-alone entry probabilities when firms ignore competition. Second, the outer one transforms an approximation of the ex-ante value of entry $\sum_{\imath=1}^K\sum_{\jmath=1}^K\beta_{\imath \jmath} s_\imath \big( \cdot \big) s_{\jmath} \big( \cdot \big)$, before observing T1EV errors, into CCPs. Altogether, our approximation structure is a hybrid of a simple neural network and a tensor product linear sieve space. It leverages the index structure in the payoff function and hence reduces the dimension of the approximation parameters needed.

\begin{rmk}
 To our knowledge, no nested fixed-point algorithm or other numerical algorithms exist for finding all equilibria in such games, rendering MLE challenging to apply. In addition, the MPEC estimator would solve a constrained maximization problem with thousands of unknowns, making it computationally difficult. Finally, NPL has no guarantee of convergence in empirical games.  
\end{rmk}

\subsection{Estimation Results}

The algorithms proposed in Section \ref{Sec:Algo} share the same asymptotic properties and perform similarly in simple settings. For practical, real-world applications, we recommend breaking the search process into more manageable steps. Specifically, we suggest using the joint algorithm with a small smoothing parameter to identify good starting values, followed by the alternating iterative version of the nested algorithm for the main estimation. Table \ref{Jia2008} shows the estimated parameters when the number of basis functions, $K$, varies from 10 to 30. The second last column reports the maximum likelihood estimates assuming equilibrium uniqueness.\footnote{We conduct fine grid search to check equilibrium uniqueness and find that the equilibrium is unique in each market.} 

The last column reports the two-step estimates. Two-step methods are generally less efficient than MLE and also rely on consistent first-stage estimates of the CCPs. Ideally, this first-stage estimation should be nonparametric, as the functional form of the solution is unknown, even when the profit and best response functions are known. However, this leads to the well-known curse of dimensionality. To address this, we propose a novel two-step approach that leverages the single-index structure, thereby avoiding the curse of dimensionality.\footnote{Similar strategies can be found in the econometrics literature on single-index regression models, such as \citet{stoker1986consistent} and \citet{powell1989semiparametric}.} Specifically, we first obtain the sieve MLE of the CCPs, $\widehat{p}_{j}$, using the same approximation structure as in Equation \eqref{oursieve}, and then estimate the parameters by maximizing the pseudo-likelihood function 
\[
\max_{\theta} \quad \sum_{j=\W,\K}\sum_{m=1}^{M}\left[d_{jm} \log\left\{p_{j}^\theta(\chi_{m})\right\} +(1-d_{jm} )\log\left\{1-p_{j}^\theta(\chi_{m})\right\}\right],
\]
where $p_{j}^\theta = \frac{1}{1+\exp \{- \xi_{j}(\theta) + \widehat{p}_{-j} \Delta\}}$.

\begin{table}[h]  \label{Jia2008}
\centering
\begin{threeparttable}
\caption{Estimation Results} 
\begin{tabular}{lr rr | r |  r} \\ \hline \hline
             & \multicolumn{3}{c}{SEES}  & \multicolumn{1}{c}{MLE}            & 2-Step \\
   K          & 10     & 20     & 30     &               &         \\ \hline 
             \textbf{Market-specific} &&&& & \\  
pop   & 3.38   & 3.37   & 3.34   & 3.29 (0.15)   & 3.40    \\
spc          & 2.81   & 2.83   & 2.82   & 2.80 (0.19)   & 2.99    \\
urban        & 2.37   & 2.39   & 2.37   & 2.32 (0.31)   & 2.50    \\
\textbf{Walmart-specific} &&&& & \\
intercept & -22.90 & -23.03 & -22.77 & -22.29 (1.73) & -24.33  \\
dbenton  & -1.86  & -1.85  & -1.87  & -1.90 (0.13)  & -1.87   \\
south     & 1.02   & 1.02   & 1.06   & 1.10 (0.15)   & 1.07    \\
\textbf{Kmart-specific} &&&& & \\
intercept & -36.21 & -36.31 & -36.22 & -35.99 (1.69) & -37.70  \\
midwest   & 0.66   & 0.66   & 0.65   & 0.65 (0.14)   & 0.58    \\ \hline 
$\Delta$     & 1.96   & 1.98   & 1.85   & 1.65 (0.27)   & 2.12   \\ \hline 
\end{tabular}
\begin{tablenotes}
\item Note: The model is estimated using the proposed method with $K = 10, 20, 30$, MLE, and the two-step method. The SEES approach results in final $\omega$ values of $10^3$, $10^4$, and $10^5$, with corresponding $\rho$ values of $0.1679$, $0.0247$, and $0.0180$, respectively.
\end{tablenotes}
\end{threeparttable}
\end{table}


The estimates are quite similar across different estimators. All estimates are significant at the 5\% level and their signs are consistent with \citet{jia2008happens}. More populated areas tend to have more stores, and higher retail sales per capita predict increased entry. Urbanized areas also attract more entry. The southern region dummy variable and the log of the distance to Walmart's headquarters in Bentonville, Arkansas, both significantly predict Walmart's entry decisions. Similarly, because Kmart’s headquarters are located in Troy, Michigan, the dummy variable for the Midwest region is predictive of Kmart's entry decisions. 

The maximum likelihood estimates imply that the mean and standard deviation of $\xi_{\W}$ are -0.11 and 3.43, respectively, while the mean and standard deviation of $\xi_{\K}$ are -2.20 and 3.30, respectively. The large coefficients on firm dummies suggest substantial entry costs. Note that Walmart is a dominant firm with a penetration rate of 48\%, while Kmart is relatively weak with a penetration rate of 19\%. This explains the much lower coefficient on the Kmart firm dummy. The proposed sieve estimator performs well across different $K$. The larger the approximation parameter $K$ is, the closer the estimates become to the maximum likelihood estimates. The estimates from the proposed two-step estimator have larger biases.


\section{Conclusion} \label{sec:conc}

A structural model is based on economic theory and describes how endogenous variables relate to a set of explanatory variables. This relationship is often expressed as an implicit function dependent on unknown parameters, which can be costly to solve. Two-step methods avoid solving the model but rely heavily on the accuracy of the first-step nonparametric estimation. We introduce \SEES\ as a new class of estimators that use a sieve to approximate the solution while penalizing deviations from the equilibrium condition. \SEES\ are straightforward to apply, at least as fast as alternative approaches like MLE, and more robust across various models. We believe our method will become a valuable tool in structural estimation.


\appendix

\newpage

\renewcommand{\thetable}{\thesection\arabic{table}} 


\section*{Appendix}

\setcounter{equation}{0}
	\setcounter{figure}{0}
	\setcounter{table}{0}
	\setcounter{theorem}{0}
	\setcounter{equation}{0}
	\makeatletter
	\renewcommand{\thetheorem}{A\arabic{theorem}}
	\renewcommand{\theequation}{A\arabic{equation}}
	\renewcommand{\theproposition}{A\arabic{proposition}}
	\renewcommand{\thefigure}{S\arabic{figure}}
	\renewcommand{\bibnumfmt}[1]{[S#1]}
	\renewcommand{\citenumfont}[1]{S#1}

\section{Optimal Monopoly Pricing With Logit Demand} \label{lambertprice}

In this subsection, we derive Equation \eqref{Plogit}. Rearranging terms gives $\xi_j - \alpha P_j + \exp(\xi_j - \alpha P_j) = \xi_j - \alpha c_j -1$, which reduces to
\[
\xi_j(x) - P_j^*(x) + \exp\{\xi_j(x) - P_j^*(x)\} = \xi_j(x) -1
\]
under our assumptions $c_j = 0,\alpha = 1$. Denote $M(x) = \exp\{\xi_j(x) - P_j^*(x)\}$. The FOC can be rewritten as $\log M(x) + M(x) = \xi_j(x) -1$. Therefore, we have 
\[
M(x) = W(\exp\{\xi_j(x) - 1\}),
\]
applying an alternative definition of the Lambert W function $\log W(v) + W(v) = \log v$. That is, the optimal price satisfies  
\begin{align*}
    P_j^*(x) = \xi_j(x) - \log M(x) = \xi_j(x) - (\xi_j(x)-1 - M(x)) = 1+W(\theta x),
\end{align*}
where the first equation follows the definition of $M(\cdot)$, the second equation follows the rewritten FOC, and the last equation follows $\xi_j(x) = \log x + \log \theta + 1$.

\section{Proofs in Section \ref{asym}}

\begin{proof}[Proof of Theorem \ref{th-pcons}]
Define $\Prob_n f(Y, p) = \frac{1}{n} \sum_{i = 1}^n f(Y_i, p)$. To continue the proof, we first show the following technical result:
\begin{equation}
\sup_{\theta \in \Theta} \Prob_n[-f(Y, \ps(\theta))] = O_{\Prob}(1). 
\label{eq-unifbounded}
\end{equation}
As Assumption \ref{ass-fixed} is met,   $\|\ps(\theta)\|_2 \leq C$ for some positive constant $C$ for any $\theta \in \Theta$. In the following proof, we assume $d_2 = 1$ to simplicity; i.e., $Y_i$ is a scalar. A vector-valued $Y_i$ can be handled in a similar way.
If $Y_i$'s are bounded, $\sup_{\theta \in \Theta} \Prob_n[-f(Y, \ps(\theta))]$ must be bounded, because $f$ is a continuous function. Hence, \eqref{eq-unifbounded} holds. If $Y_i$'s are not bounded, by Assumption \ref{ass-bounded}, there exists a positive constant $\eta$ such that
$$
\eta \sup_{x \in K} [-f(y, x)] \leq 1 + \inf_{x \in K} [-f(y, x)]\quad\forall y \in \Re.
$$
Therefore, 
$$
\sup_{\theta \in \Theta} \Prob_n[-f(Y, \ps(\theta))] \leq \eta^{-1} \{1 + \Prob_n [-f(Y, \ps(\theta_0))]\}.
$$
By the strong law of large numbers, with probability one, we have 
$$
\Prob_n [-f(Y, \ps(\theta_0))] \rightarrow \E_{\theta_0}[-f(Y, \ps(\theta_0))] = -M(\theta_0) < \infty.
$$
It follows that $\eta^{-1} \{1 + \Prob_n [-f(Y, p(\theta_0))]\} = O_{\Prob}(1)$. Equation \eqref{eq-unifbounded} is established.

Let $J(p, \theta) = \|p - \Psi(p, \theta)\|_2^2$. 
Based on the definition of $\hat{p}(\theta)$, we have that, for any $\theta \in \Theta$, 
\begin{align*}
    \ell_n(\hat{p}(\theta)) - \omega_n J(\hat{p}(\theta), \theta) \geq \ell_n(\ps(\theta)).
\end{align*}
As $\ell_n(\hat{p}(\theta)) \leq 0$ by Assumption \ref{ass-bounded}, it follows that
$$
- \omega_n J(\hat{p}(\theta), \theta) \geq \ell_n(\ps(\theta)).
$$
Then, 
\begin{align*}
    J(\hat{p}(\theta), \theta) & \leq -\omega_n^{-1} \ell_n(\ps(\theta))  =  \omega_n^{-1} \Prob_n[-f(Y, \ps(\theta))].
    \end{align*}
Therefore, it follows from \eqref{eq-unifbounded} that
$\sup_{\theta \in \Theta}  J(\hat{p}(\theta), \theta)
  = \omega_n^{-1} O_{\Prob}(1)$.
Consequently, 
$$
\sup_{\theta \in \Theta}\|g(\hat{p}(\theta), \theta)\|_2 = O_{\Prob}\left(\frac{1}{\rt{\omega_n}}\right),
$$ while $g(\ps(\theta), \theta) = 0$. 
By Assumption \ref{ass-Lip}, we have
$$
\sup_{\theta \in \Theta}  \|\hat{p}(\theta) - \ps(\theta)\|_2 \leq O_{\Prob}\left(\frac{1}{\sqrt{\omega_n}}\right). 
$$
This completes the proof. 
\end{proof}

\begin{proof}[Proof of Theorem \ref{th-jointcons}]
We first show that
\begin{equation} \label{eq-error}
    \sup_{\theta \in \Theta}|\ell_n(\hat{p}(\theta)) - \ell_n(\ps(\theta))| = O_{\Prob}\left(\frac{1}{\sqrt{\omega_n}}\right) + o_{\Prob}\left(\frac{1}{n}\right).
\end{equation}

Recall that $M(\theta) = \E_{\theta_0}[f(Y_i, \ps(\theta))]$. Then, we define
$$
M_n(\theta) = \ell_n(\ps(\theta)) = \frac{1}{n}\sum_{i = 1}^n f(Y_i, \ps(\theta)) \quad \text{for}~\theta \in \Theta.
$$
We will show 
\begin{equation} \label{eq-unifLLN}
    \sup_{\theta \in \Theta}|M_n(\theta) - M(\theta)| = o_{\Prob}(1).
\end{equation}

Finally, we prove that 
$$
\tilde{\theta}_n \rightarrow \theta_0
$$
in probability as $n \rightarrow \infty$. 

To prove \eqref{eq-error}, we assume that $\Theta$ is convex without loss of generality. By Assumption \ref{ass-fixed}, there must exist some positive constant $r$ such that
$$
\|\ps(\theta)\|_2 \leq r,\quad\quad \forall \theta \in \Theta.  
$$
Let $V_n = \sup_{\theta \in \Theta} \|\hat{p}(\theta) - \ps(\theta)\|_{2}$.
Theorem \ref{th-pcons} indicates 
$$
V_n \leq  O_{\Prob}\left(\frac{1}{\sqrt{\omega_n}}\right). 
$$
By Assumption \ref{ass-strong}, there exists a positive constant $\eta$ such that, for $j = 1, \ldots, d_1$,
\begin{equation} \label{eq-eta}
\sup_{\|p\|_2 \leq r + 1} \left|\frac{\p f}{\p p_j}(y, p) \right|
\leq \frac{1}{\eta} \left\{1 + \inf_{\|p\|_2 \leq r + 1} \left|\frac{\p f}{\p p_j}(y, p) \right|\right\} \quad \forall y \in \Re. 
\end{equation}
Then, it follows that 
\begin{align*}
    &\quad\quad |\ell_n(\hat{p}(\theta)) - \ell_n(\ps(\theta))| \\
    & \leq \frac{1}{n} \sum_{i = 1}^n |f(Y_i,\hat{p}(\theta)) -
    f(Y_i, \ps(\theta))| \\
    & = \frac{1}{n} \sum_{i = 1}^n |f(Y_i,\hat{p}(\theta)) -
    f(Y_i, \ps(\theta))|\mathbbm{1}_{(V_n \leq 1)} +  
    \frac{1}{n} \sum_{i = 1}^n |f(Y_i,\hat{p}(\theta)) -
    f(Y_i, \ps(\theta))|\mathbbm{1}_{(V_n > 1)}\\
    & \leq \frac{1}{n} \sum_{i = 1}^n \left[\sup_{\substack{\|p\|_2 \leq r + 1, \\1 \leq j \leq d_1}} \left|\frac{\p f}{\p p_j}(Y_i, p)\right|\right] \|\hat{p}(\theta) - \ps(\theta)\|_{2}\mathbbm{1}_{(V_n \leq 1)}\\ 
    & \qquad\qquad\qquad  + \frac{1}{n} \sum_{i = 1}^n |f(Y_i,\hat{p}(\theta)) -
    f(Y_i, \ps(\theta))|\mathbbm{1}_{(V_n > 1)}\\
    & \leq \frac{1}{n} \sum_{i = 1}^n \frac{1}{\eta} \left\{1 + \max_{1 \leq j \leq d_1}\inf_{\|p\|_2 \leq r + 1} \left|\frac{\p f}{\p p_j}(Y_i, p) \right|\right\} V_n \mathbbm{1}_{(V_n \leq 1)}  \\
    & \qquad\qquad\qquad + \frac{1}{n} \sum_{i = 1}^n |f(Y_i,\hat{p}(\theta)) -
    f(Y_i, \ps(\theta))|\mathbbm{1}_{(V_n > 1)}\\
    & \leq \frac{1}{n} \sum_{i = 1}^n \frac{1}{\eta} \left\{1 +  \left\|\frac{\p f}{\p p}(Y_i, \ps(\theta_0)) \right\|_2\right\} V_n
     + \frac{1}{n} \sum_{i = 1}^n |f(Y_i,\hat{p}(\theta)) -
    f(Y_i, \ps(\theta))|\mathbbm{1}_{(V_n > 1)}. 
\end{align*}
Therefore, 
\begin{align} \label{eq-maxbound}
\begin{split}
    \sup_{\theta \in \Theta}|\ell_n(\hat{p}(\theta)) - \ell_n(\ps(\theta))|  \leq & \frac{1}{n} \sum_{i = 1}^n \frac{1}{\eta} \left\{1 +  \left\|\frac{\p f}{\p p}(Y_i, \ps(\theta_0)) \right\|_2\right\} V_n \\
     & \quad\quad + \frac{1}{n} \sum_{i = 1}^n |f(Y_i,\hat{p}(\theta)) -
    f(Y_i, \ps(\theta))|\mathbbm{1}_{(V_n > 1)}. 
\end{split}
\end{align}
By the strong law of large numbers and Assumption \ref{ass-strong}, 
$$
\frac{1}{n} \sum_{i = 1}^n \left[1 +  \left\|\frac{\p f}{\p p}(Y_i, \ps(\theta_0)) \right\|_2 \right] \rightarrow 1 + \E_{\theta_0} \left\|\frac{\p f}{\p p}(Y_i, \ps(\theta_0)) \right\|_2
$$
almost surely. So, it is $O_{\Prob}(1)$. The second term on the right-hand side of \eqref{eq-maxbound} is nonzero only in the event $\{V_n > 1\}$, whose probability converges to zero by Theorem \ref{th-pcons}, so it is $o_{\Prob}(n^{-1})$. Hence, we have established Equation \eqref{eq-error}.
Equation \eqref{eq-unifLLN} follows from Lemma \ref{le-unifSLLN}, which will be presented later. 

Now, we are ready to prove $\tilde{\theta}_n \rightarrow \theta_0$ in probability. Let $\delta$ be an arbitrary positive number. Assumption \ref{ass-unimax} indicates that $\theta_0$ is the unique maximizer of 
$M(\theta)$. As $M(\theta)$ is continuous over the compact set $\Theta$, 
\begin{equation} \label{eq-Mest}
    \gamma := M(\theta_0) - \sup_{\theta \in \Theta, d(\theta, \theta_0) \geq \delta}M(\theta) > 0,
\end{equation}
where $d(\theta, \theta_0) = \|\theta - \theta_0\|_2$ for any $\theta \in \Theta$. Note that $\tilde{\theta}_n \in \Theta$. By \eqref{eq-error} and \eqref{eq-unifLLN}, we have
\begin{align*}
    & \quad\quad M(\theta_0) - M(\tilde{\theta}_n) \\
    & = M_n(\theta_0) - M_n(\tilde{\theta}_n) + o_{\Prob}(1) \\
    & = \ell_n(\ps(\theta_0)) - \ell_n(\ps(\tilde{\theta}_n)) + o_{\Prob}(1)\\
    & \leq  \ell_n(\ps(\theta_0)) - \ell_n(\hat{p}(\tilde{\theta}_n)) + \sup_{\theta \in \Theta}|\ell_n(\hat{p}(\theta)) - \ell_n(\ps(\theta))| + o_{\Prob}(1) \\
  & = \ell_n(\ps(\theta_0)) - \ell_n(\hat{p}(\tilde{\theta}_n)) + o_{\Prob}(1).
\end{align*}
By definition of $\tilde{\theta}_n$ and $\ps(\theta_0)$, we have
$$
\ell_n(\hat{p}(\tilde{\theta}_n)) \geq \ell_n(\ps(\theta_0)),
$$
so $M(\theta_0) - M(\tilde{\theta}_n) \leq o_{\Prob}(1)$. By \eqref{eq-Mest}, one has
$$
\{d(\tilde{\theta}_n, \theta_0) \geq \delta\} \subset \{M(\theta_0) - M(\tilde{\theta}_n) \geq \gamma\} \subset \{o_{\Prob}(1) \geq \gamma\}.
$$
Therefore, 
$$
\Prob_{\theta_0}(d(\tilde{\theta}_n, \theta_0) \geq \delta) \leq \Prob_{\theta_0} (o_{\Prob}(1) \geq \gamma),
$$
which converges to 0 as $n \rightarrow \infty$. As $\delta$ is an arbitrary positive number, $\tilde{\theta}_n$ is a consistent estimator of $\theta_0$. This completes the proof. 
\end{proof}

\begin{proof}[Proof of Theorem \ref{th-jointasynormal}]
We mainly follow Theorem 5.23 of \cite{van2000} to prove asymptotic normality of $\tilde{\theta}_n$.
Firstly, as we have shown in the proof of Lemma \ref{le-unifSLLN}, 
\begin{align}
    \nonumber
    &\quad\quad |f(y, \ps(\theta_1)) - f(y, \ps(\theta_2))| \\
    \nonumber
    & \leq \frac{1}{\eta} \left[1 +  \left\|\frac{\p f}{\p p}(y, \ps(\theta_0)) \right\|_2\right] \|\ps(\theta_1) - \ps(\theta_2)\|_2 \\
    & \leq \frac{C}{\eta} \left[1 +  \left\|\frac{\p f}{\p p}(y, \ps(\theta_0)) \right\|_2\right] \|\theta_1 - \theta_2\|_2
    \label{eq-equicontinuous}
\end{align}
for some constant $C$, and $\eta$ is defined in Equation \eqref{eq-eta}. By Assumption \ref{ass-normal}, the right-hand side of \eqref{eq-equicontinuous} has a finite second moment.

Next, we consider a second-order Tayor expansion for
$$
M(\theta) = \E_{\theta_0}[f(Y, \ps(\theta))]
$$
in a neighbourhood of $\theta_0$. Obviously, 
\begin{align}
\nonumber
&\quad\quad f(y, p(\theta)) \\
\nonumber 
& = f(y, \ps(\theta_0)) \\
\label{eq-linterm}
&\quad+ \left[\frac{\p f(y, \ps(\theta_0))}{\p p}\right]^{\prime}\left(\frac{\p  \ps(\theta_0)}{\p \theta}\right) (\theta - \theta_0) \\
\nonumber
&\quad + \frac{1}{2} (\theta-\theta_0)^{\prime} \left[\sum_{j = 1}^{d_1} \frac{\p f(y, \ps(\theta_0))}{\p p_j}
\frac{\p^2  \ps_j(\theta_0)}{\p \theta \p \theta^{\prime}} + \left(\frac{\p  \ps(\theta_0)}{\p \theta}\right)^{\prime} \frac{\p^2 f(y, \ps(\theta_0))}{\p p \p p^{\prime}}\left(\frac{\p  \ps(\theta_0)}{\p \theta}\right) \right]  \\
\nonumber
& \quad \times (\theta - \theta_0) + R,
\end{align}
where $R$ is the remainder term. Define
$$
D(y, \theta) = \sum_{j = 1}^{d_1} \frac{\p f(y, \ps(\theta))}{\p p_j}
\frac{\p^2  \ps_j(\theta)}{\p \theta \p \theta^{\prime}} + \left(\frac{\p  \ps(\theta)}{\p \theta}\right)^{\prime} \frac{\p^2 f(y, \ps(\theta))}{\p p \p p^{\prime}}\left(\frac{\p  \ps(\theta)}{\p \theta}\right).
$$
Then the reminder term can be rewritten as
$$
R = (\theta - \theta_0)^{\prime} \left[\int_0^1 [D(y, \theta_0 + s(\theta - \theta_0)) - D(y, \theta_0)] (1 - s) \diff s \right] (\theta - \theta_0).
$$
Note that $D(y, \theta)$ is a $p \times p$ matrix. For any $(a,b)$th entry in $D(y, \theta)$, by using the same argument for Equation \eqref{eq-maxbound},
we can show that, for any $\theta \in \Theta$,
\begin{align*}
 & \quad\quad  |D_{ab}(y, \theta)| \\
 & \leq \sum_{j = 1}^{d_1} \left|\frac{\p f(y, \ps(\theta))}{\p p_j}
\frac{\p^2  \ps_j(\theta)}{\p \theta_a \p \theta_b}\right| + \left|\left(\frac{\p  \ps(\theta)}{\p \theta_a}\right)^{\prime} \frac{\p^2 f(y, \ps(\theta))}{\p p \p p^{\prime}}\left(\frac{\p  \ps(\theta)}{\p \theta_b}\right)\right| \\
& \leq \frac{C^{\prime}}{\eta} \left[1 +  \left\|\frac{\p f}{\p p}(y, \ps(\theta_0)) \right\|_2 + \max_{i,j}\left|\frac{\p^2 f}{\p p_i \p p_j}(y, \ps(\theta_0))\right|\right],
\end{align*}
where $C^\prime$ is a positive constant. Additionally, under Assumption \ref{ass-normal}, the right-hand side of the above inequality has a finite mean. Therefore, applying the dominated convergence theorem, we have
$$
\E_{\theta_0} \left[\int_0^1 [D(y, \theta_0 + s(\theta - \theta_0)) - D(y, \theta_0)] (1 - s) \diff s \right] \rightarrow 0
$$
as $\theta \rightarrow \theta_0$. Then, by the Taylor expansion of $f(y, \ps(\theta))$, it follows that 
\begin{equation} \label{eq-Taylor}
    M(\theta) = M(\theta_0) + \frac{1}{2} (\theta - \theta_0)^{\prime} V_{\theta_0} (\theta - \theta_0) + o\left(\|\theta - \theta_0\|_2^2\right).
\end{equation}
Recall that $f(y, p)$ is the log density of $Y_i$.
Therefore, there is no linear form in \eqref{eq-Taylor} and the expected value of $D(Y, \theta_0)$ is given by $V_{\theta_0}$
as the expectation of \eqref{eq-linterm} is zero.

Finally, we want to establish that
\begin{equation} \label{eq-nearmaxjoint}
    \ell_n(\ps(\tilde{\theta}_n)) \geq \sup_{\theta \in \Theta} \ell_n(\ps(\theta)) - o_{\Prob}(n^{-1}). 
\end{equation}
By definition of $\ps(\theta)$ and $\tilde{\theta}_n$, we have
\begin{align*}
&\quad\quad \ell_n(\ps(\tilde{\theta}_n)) \\
& \geq \ell_n(\hat{p}(\tilde{\theta}_n)) - \sup_{\theta \in \Theta} |\ell_n(\hat{p}(\theta)) - \ell_n(\ps(\theta))| \\
& \geq \sup_{\theta \in \Theta} \ell_n(\ps(\theta))
- \sup_{\theta \in \Theta} |\ell_n(\hat{p}(\theta)) - \ell_n(\ps(\theta))|. 
\end{align*}
By \eqref{eq-error} and $\omega_n/n^2 \rightarrow \infty$, 
we have
$$
\sup_{\theta \in \Theta}|\ell_n(\hat{p}(\theta)) - \ell_n(\ps(\theta))| = O_{\Prob}\left(\frac{1}{\sqrt{\omega_n}}\right) + o_{\Prob}\left(\frac{1}{n}\right) = o_{\Prob}(n^{-1}). 
$$
Hence, the relation in \eqref{eq-nearmaxjoint} holds. 
By \eqref{eq-equicontinuous}, \eqref{eq-Taylor}, \eqref{eq-nearmaxjoint}, and Theorem \ref{th-jointcons}, it follows from Theorem 5.23 in \cite{van2000} that $\sqrt{n}(\tilde{\theta}_n - \theta_0)$ is asymptotically normal with mean zero and covariance matrix
$$
V_{\theta_0}^{-1} \left(\frac{\p  \ps(\theta_0)}{\p \theta}\right)^{\prime} \E_{\theta_0} \left[\left\{\frac{\p f(y, \ps(\theta_0))}{\p p}\right\}\left\{\frac{\p f(y, \ps(\theta_0))}{\p p}\right\}^{\prime}\right]\left(\frac{\p  \ps(\theta_0)}{\p \theta}\right) V_{\theta_0}^{-1},
$$
if $V_{\theta_0}$ is non-singular. 
This completes the proof. 
\end{proof}

\begin{lemma} \label{le-unifSLLN}
The class $\{f(\cdot, \ps(\theta)), \theta \in \Theta\}$ is $\Prob_{\theta_0}$-Glivenko-Cantelli. 
\end{lemma}

\begin{proof} 
As Assumption \ref{ass-ident} is met and $\Theta$ is a compact set, by the implicit function theorem, there exists some constant $C$ such that 
\begin{align*}
\|\ps(\theta_1) - \ps(\theta_2) \|_2 \leq C \|\theta_1 - \theta_2\|_2
\end{align*}
for any $\theta_1, \theta_2 \in \Theta$. Furthermore, with a similar argument for Equation \eqref{eq-maxbound}, we obtain
\begin{align}
    \nonumber
    &\quad\quad |f(y, \ps(\theta_1)) - f(y, \ps(\theta_2))| \\
    \nonumber
    & \leq \frac{1}{\eta} \left[1 +  \left\|\frac{\p f}{\p p}(y, \ps(\theta_0)) \right\|_2\right] \|\ps(\theta_1) - \ps(\theta_2)\|_2 \\
    & \leq \frac{C}{\eta} \left[1 +  \left\|\frac{\p f}{\p p}(y, \ps(\theta_0)) \right\|_2\right] \|\theta_1 - \theta_2\|_2.
\end{align}

By Assumption \ref{ass-strong}, $\left\|\frac{\p f}{\p p}(Y, \ps(\theta_0)) \right\|_2$ has a finite expectation under $\Prob_{\theta_0}$. Thus, based on Theorem 2.7.11 in \cite{van1996}, the $L_1(\Prob_{\theta_0})$-bracketing number is bounded by the covering number $N(\epsilon, \Theta_0, \|\cdot\|_2)$ of $\Theta_0$. Since $\Theta_0$ is a compact subset of $\Re^d$, 
$$
N(\epsilon, \Theta, \|\cdot\|) \leq C_0 \times \left(\frac{1}{\epsilon}\right)^{d}
$$
for some constant $C_0$ and any $\epsilon > 0$.
Therefore, by Theorem 2.4.1 of \cite{van1996}, this lemma holds.
\end{proof}

\begin{proof}[Proof of Corollary \ref{fisherjoint}]
Consider the following problem:
$\max\quad h(\beta,\theta)=\ell(\beta,\theta)- \omega\left(\Psi(\beta,\theta)-\beta\right)^{2}$
where $\beta=p$. Taking the first-order condition gives
\begin{align*}
\ell_{\beta} - 2\omega\left(\Psi-\beta\right)\left(\Psi_{\beta}-1\right) & =0\\
\ell_{\theta} - 2\omega\left(\Psi-\beta\right)\Psi_{\theta} & =0
\end{align*}
Note that $\Psi(\widehat{\beta}(\theta),\theta)-\widehat{\beta}(\theta) \approx 0$ for  $\omega$ approaches infinity, which further implies that 
$\ell_{\beta} \approx 0$. Taking the derivative gives 
$\Psi_{\beta}\widehat{\beta}^{\prime}(\theta)+\Psi_{\theta}-\widehat{\beta}^{\prime}(\theta) \approx 0$,
which implies that 
$\Psi_{\beta}-1 \approx -\frac{\Psi_{\theta}}{\widehat{\beta}^{\prime}(\theta)}$.

Taking the second-order derivative gives
$\left(\begin{array}{cc}
\frac{\partial^{2}h}{\partial\theta\partial\theta} & \frac{\partial^{2}h}{\partial\theta\partial\beta}\\
\frac{\partial^{2}h}{\partial\theta\partial\beta} & \frac{\partial^{2}h}{\partial\beta\partial\beta}
\end{array}\right)$, where 
\begin{align*}
\frac{\partial^{2}h}{\partial\theta\partial\theta} & =\ell_{\theta\theta} - 2\omega \left[\left(\Psi_{\theta}\right)^{2}+\left(\Psi-\beta\right)\Psi_{\theta\theta}\right]\\
\frac{\partial^{2}h}{\partial\theta\partial\beta} & =\ell_{\theta\beta} - 2\omega \left[\left(\Psi_{\beta}-1\right)\Psi_{\theta}+\left(\Psi-\beta\right)\Psi_{\beta\theta}\right]\\
\frac{\partial^{2}h}{\partial\beta\partial\beta} & =\ell_{\beta\beta} -2\omega \left[\left(\Psi_{\beta}-1\right)^{2}+\left(\Psi-\beta\right)\Psi_{\beta\beta}\right]
\end{align*}

As $\omega$ approaches infinity, we study the block that we highlight here\footnote{Note that the inverse of a block matrix 
$\left(\begin{array}{cc}
A & B\\
C & D
\end{array}\right)^{-1}=\left(\begin{array}{cc}
\left(A-BD^{-1}C\right)^{-1} & *\\
* & *
\end{array}\right)$.}
\begin{align*}
 & \bf{H}_{\theta \theta} - \bf{H}_{\beta \theta}^\prime \bf{H}_{\beta \beta}^{-1} \bf{H}_{\beta \theta} \\
= & \ell_{\theta\theta} - 2\omega \left(\Psi_{\theta}\right)^{2}-\frac{\left[\ell_{\theta\beta}+2\omega \frac{\left(\Psi_{\theta}\right)^{2}}{\widehat{\beta}^{\prime}(\theta)}\right]^{2}}{\ell_{\beta\beta}-2\omega \left(\frac{\Psi_{\theta}}{\widehat{\beta}^{\prime}(\theta)}\right)^{2}}\\
= & \ell_{\theta\theta}- \frac{2\omega\left(\Psi_{\theta}\right)^{2}\ell_{\beta\beta}-4\omega^2\left(\Psi_{\theta}\right)^{2}\left(\frac{\Psi_{\theta}}{\widehat{\beta}^{\prime}(\theta)}\right)^{2}+\left(\ell_{\theta\beta}\right)^{2}+4\omega^2\frac{\left(\Psi_{\theta}\right)^{4}}{(\widehat{\beta}^{\prime}(\theta))^2}+4\omega\ell_{\theta\beta}\frac{\left(\Psi_{\theta}\right)^{2}}{\widehat{\beta}^{\prime}(\theta)}}{\ell_{\beta\beta}-2\omega\left(\frac{\Psi_{\theta}}{\widehat{\beta}^{\prime}(\theta)}\right)^{2}}\\
\rightarrow & \ell_{\theta\theta} + \ell_{\beta\beta}\left[\widehat{\beta}^{\prime}(\theta)\right]^{2} + 2\ell_{\theta\beta}\widehat{\beta}^{\prime}(\theta)
\end{align*}

Now consider $\ell(\widehat{\beta}(\theta),\theta)$, where 
$\widehat{\beta}(\theta)$ solves $\Psi(\beta,\theta)=\beta$. We have the Hessian 
\[
\ell_{\theta\theta}+\ell_{\beta\beta}\left[\widehat{\beta}^{\prime}(\theta)\right]^{2}+2\ell_{\theta\beta}\widehat{\beta}^{\prime}(\theta)+\ell_{\beta}\widehat{\beta}^{\prime\prime}(\theta)=\ell_{\theta\theta}+\ell_{\beta\beta}\left[\widehat{\beta}^{\prime}(\theta)\right]^{2}+2\ell_{\theta\beta}\widehat{\beta}^{\prime}(\theta),
\]
which equals the limit of $\bf{H}_{\theta \theta} - \bf{H}_{\beta \theta}^\prime \bf{H}_{\beta \beta}^{-1} \bf{H}_{\beta \theta}$.
\end{proof}

\begin{proof}[Proof of Corollary \ref{MPEChessian}]
Denote the $j$th function of $g(\widehat{\beta}(\theta),\theta)=0$ as $g^j$. Taking its first-order derivative gives
\[
\sum_{k}g_{\beta_{k}}^{j}\frac{\partial\beta_{k}}{\partial\theta_{\ell}}+g_{\theta_{\ell}}^{j} =0 , 
\]
which can be written in matrix form $\nabla_\beta g \times  \nabla\beta(\theta)+\nabla_\theta g=0$. Therefore,
\[
\nabla\beta(\theta)=-\left[\nabla_\beta g\right]^{-1} \times \nabla_\theta g .
\]
Taking its second-order derivative gives 
\begin{align} \label{gtheta2}
\sum_{k}\left\{ \left[\sum_{k^{\prime}}g_{\beta_{k}\beta_{k^{\prime}}}^{j}\frac{\partial\beta_{k^{\prime}}}{\partial\theta_{\ell^{\prime}}}+g_{\beta_{k}\theta_{\ell^{\prime}}}^{j}\right]\frac{\partial\beta_{k}}{\partial\theta_{\ell}}+g_{\beta_{k}\theta_{\ell}}^{j}\frac{\partial\beta_{k}}{\partial\theta_{\ell^{\prime}}}\right\} +\sum_{k}g_{\beta_{k}}^{j}\frac{\partial^{2}\beta_{k}}{\partial\theta_{\ell}\partial\theta_{\ell^{\prime}}}+g_{\theta_{\ell}\theta_{\ell^{\prime}}}^{j} & =0.
\end{align}

Now, consider the Hessian of the likelihood function $\widehat{\ell}(\theta) = \ell(\widehat{\beta}(\theta), \theta)$, 

\begin{align*}
\begin{split}
\frac{\partial^2 \widehat{\ell}(\theta)}{\partial \theta_\ell \theta_{\ell^\prime}}
= &  \sum_{k}\left\{ \left[\sum_{k^{\prime}}\ell_{\beta_{k}\beta_{k^{\prime}}}\frac{\partial\beta_{k^{\prime}}}{\partial\theta_{\ell^{\prime}}}+\ell_{\beta_{k}\theta_{\ell^{\prime}}}\right]\frac{\partial\beta_{k}}{\partial\theta_{\ell}}+\ell_{\beta_{k}}\frac{\partial^{2}\beta_{k}}{\partial\theta_{\ell}\partial\theta_{\ell^{\prime}}}+\ell_{\beta_{k}\theta_{\ell}}\frac{\partial\beta_{k}}{\partial\theta_{\ell^{\prime}}}\right\} +\ell_{\theta_{\ell}\theta_{\ell^{\prime}}} \\
= &
\ell_{\theta_{\ell}\theta_{\ell^{\prime}}} \!+\!\sum_{k}\sum_{k^{\prime}}\ell_{\beta_{k}\beta_{k^{\prime}}}\frac{\partial\beta_{k^{\prime}}}{\partial\theta_{\ell^{\prime}}}\frac{\partial\beta_{k}}{\partial\theta_{\ell}} \!+\! \sum_{k}\left[\ell_{\beta_{k}\theta_{\ell^{\prime}}}\frac{\partial\beta_{k}}{\partial\theta_{\ell}}+\ell_{\beta_{k}\theta_{\ell}}\frac{\partial\beta_{k}}{\partial\theta_{\ell^{\prime}}}\right] \\
&  \qquad \quad - \sum_{k}\sum_{j}\lambda_{j}g_{\beta_{k}}^{j}\frac{\partial^{2}\beta_{k}}{\partial\theta_{\ell}\partial\theta_{\ell^{\prime}}} \\
= & 
\underbrace{ \left[\ell_{\theta_{\ell}\theta_{\ell^{\prime}}}+\sum_{j}\lambda_{j}g_{\theta_{\ell}\theta_{\ell^{\prime}}}^{j}\right]}_{\bf{H}_{\theta \theta}}
+
\sum_{k}\sum_{k^{\prime}}
\underbrace{\left[\ell_{\beta_{k}\beta_{k^{\prime}}}+\sum_{j}\lambda_{j}g_{\beta_{k}\beta_{k^{\prime}}}^{j}\right]}_{\bf{H}_{\beta\beta }}
\frac{\partial\beta_{k^{\prime}}}{\partial\theta_{\ell^{\prime}}}\frac{\partial\beta_{k}}{\partial\theta_{\ell}} \\
& +\sum_{k} 
\underbrace{\left[\ell_{\beta_{k}\theta_{\ell^{\prime}}}+\sum_{j}\lambda_{j}g_{\beta_{k}\theta_{\ell^{\prime}}}^{j}\right]}_{\bf{H}_{\beta \theta}}
\frac{\partial\beta_{k}}{\partial\theta_{\ell}}+
\sum_{k}
\underbrace{\left[\ell_{\beta_{k}\theta_{\ell}}+\sum_{j}\lambda_{j}g_{\beta_{k}\theta_{\ell}}^{j}\right]}_{\bf{H}_{\theta \beta}}
\frac{\partial\beta_{k}}{\partial\theta_{\ell^{\prime}}} 
\end{split}
\end{align*}
where the first equation follows from the Lagrange multiplier theorem, i.e., $\ell_{\beta_k} +\sum_j \lambda_j g_{\beta_k}^j = 0$, and the second equation follows from \eqref{gtheta2}. In its matrix form, we can construct the observed Fisher information,
\[
\bf{\widehat{H}} = \bf{H}_{\theta \theta} 
+ [\nabla \widehat{\beta}(\widehat{\theta})]^\prime 
\bf{H}_{\beta \beta}[\nabla \widehat{\beta}(\widehat{\theta})] +
[\nabla \widehat{\beta}(\widehat{\theta})]^\prime \bf{H}_{\beta \theta}  + 
[\bf{H}_{\beta \theta}]^\prime \nabla \widehat{\beta}(\widehat{\theta}) , 
\]
where $\nabla\beta(\widehat{\theta})=-\left[\nabla_\beta g\right]^{-1} \times \nabla_\theta g$, and the matrices in bold are the four blocks in the Hessian matrix generated from a constrained maximization algorithm, 
\begin{align*}
        \begin{bmatrix}
\bf{H}_{\beta \beta}  & \bf{H}_{\beta \theta} \\
\bf{H}_{\theta \beta}  & \bf{H}_{\theta \theta} 
\end{bmatrix} .
\end{align*}
\end{proof}

\section{More about the Nested Estimator} \label{sec:nested}

This section provides more details on the nested estimator. First, we derive the gradient of the outer loop problem. Second, we establish its asymptotic properties. 

\begin{proposition} \label{prop-grad}
The gradient of the outer loop satisfies
\[
\nabla \widehat{\ell}(\theta) = \nabla_\theta \ell(\widehat{\beta}(\theta),\theta)  + (\nabla \widehat{\beta}(\theta))^{\prime} \times [\nabla_\beta \ell(\widehat{\beta}(\theta), \theta)]. 
\]
\end{proposition}

\begin{proof}[Proof of Proposition \ref{prop-grad}]

Fix the smoothing parameter $\omega$ for the moment. Recall that $\widehat{\beta}(\theta)$ is defined implicitly by an equation system 
\[ \frac{\partial \ell(\beta,\theta)}{\partial \beta_k}  - \omega \frac{\partial \rho(\beta,\theta)}{\partial \beta_k} = 0 . \]
Denote this system as $h_k(\beta, \theta) =0$, where $k=1,\ldots, K$. Taking the derivative (w.r.t. $\theta$) gives $\sum_\ell \frac{\partial h_k}{\partial \beta_\ell} \frac{d \beta_\ell}{d \theta} + \frac{\partial h_k}{\partial \theta} =0$. 
It allows us to find $\nabla \widehat{\beta}(\theta)$ using the Hessian of $h$ with respect to $\beta$, i.e., $\bf{H}_{\beta \beta}(\theta)$, and the cross derivative of $h$ with respect to $\beta$ and $\theta$, i.e. $\bf{H}_{\beta \theta}(\theta)$, as follows:
\[
\nabla \widehat{\beta}(\theta) = - \bf{H}_{\beta \beta}(\theta)^{-1} \bf{H}_{\beta \theta}(\theta) ,
\]
where the terms on the RHS are calculated at the inner loop solution $\beta = \widehat{\beta}(\theta)$. Therefore, the gradient of $\ell(p^{\widehat{\beta}(\theta)})$ w.r.t. $\theta$ can be calculated as follows 
\[\frac{\partial \widehat{\ell}(\theta)}{\partial \theta_\ell} 
= \frac{\partial \ell(\beta, \theta)}{\partial \theta_\ell} \bigg|_{\beta = \widehat{\beta}(\theta)} 
+ \sum_{k=1}^K  \frac{\partial \ell(\beta, \theta)}{\partial \beta_k} \bigg|_{\beta = \widehat{\beta}(\theta)} \frac{\partial \widehat{\beta}_k(\theta)}{\partial \theta_\ell},
\]
which can be written in matrix form. 
\end{proof}

\subsubsection*{Asymptotic Properties for the Nested Estimator}

Recall that $\hat{\theta}_n$ obtained from the nested algorithm is the maximizer of $\ell_n(\hat{p}(\theta))$. Similar to the joint estimator, we now establish consistency for $\tilde{\theta}_n$ as an estimator of $\theta_0$. 

\begin{theorem} \label{th-thetacons}
Under conditions of Theorem \ref{th-jointcons}, $\hat{\theta}_n$ is consistent. 
\end{theorem}


\begin{proof}[Proof of Theorem \ref{th-thetacons}]
 We employ the techniques of proving Theorem \ref{th-jointcons}.
 By \eqref{eq-error} and \eqref{eq-unifLLN}, we have
  \begin{align*}
    & \quad\quad M(\theta_0) - M(\hat{\theta}_n) \\
    & = M_n(\theta_0) - M_n(\hat{\theta}_n) + o_{\Prob}(1)\quad\text{by}~\eqref{eq-unifLLN} \\
    & = \ell_n(\ps(\theta_0)) - \ell_n(\ps(\hat{\theta}_n)) + o_{\Prob}(1)\\
    & \leq \ell_n(\hat{p}(\theta_0)) + \sup_{\theta \in \Theta}|\ell_n(\hat{p}(\theta)) - \ell_n(\ps(\theta))| - \ell_n(\hat{p}(\hat{\theta}_n)) \\
    & \quad \quad + \sup_{\theta \in \Theta}|\ell_n(\hat{p}(\theta)) - \ell_n(\ps(\theta))| + o_{\Prob}(1) \\
    & = \ell_n(\hat{p}(\theta_0)) - \ell_n(\hat{p}(\hat{\theta}_n)) 
    + 2\sup_{\theta \in \Theta}|\ell_n(\hat{p}(\theta)) - \ell_n(\ps(\theta))| + o_{\Prob}(1) \\
    & \leq \ell_n(\hat{p}(\theta_0)) - \ell_n(\hat{p}(\hat{\theta}_n)) + o_{\Prob}(1).
\end{align*}
By definition of $\hat{\theta}_n$, we have
$$
\ell_n(\hat{p}(\hat{\theta}_n)) \geq \ell_n(\hat{p}(\theta_0)),
$$
so $M(\theta_0) - M(\hat{\theta}_n) \leq o_{\Prob}(1)$. By \eqref{eq-Mest}, we have
$$
\{d(\hat{\theta}_n, \theta_0) \geq \delta\} \subset \{M(\theta_0) - M(\hat{\theta}_n) \geq \gamma\} \subset \{o_{\Prob}(1) \geq \gamma\}.
$$
Therefore, 
$$
\Prob_{\theta_0}(d(\hat{\theta}_n, \theta_0) \geq \delta) \leq \Prob_{\theta_0} (o_{\Prob}(1) \geq \gamma),
$$
which converges to 0 as $n$ approaches infinity. As $\delta$ is an arbitrary positive number, $\hat{\theta}_n$ is a consistent estimator of $\theta_0$. This completes the proof. \end{proof}

The following theorem indicates the nested estimator has the same asymptotic distribution as the joint estimator under mild conditions.

\begin{theorem} \label{th-asynormal}
Under conditions of Theorem \ref{th-jointasynormal}, $$
\rt{n}(\hat{\theta}_n - \theta_0) \overset{d}{\to}
\mN(0, \Sigma).
$$
\end{theorem}

\begin{proof}[Proof of Theorem \ref{th-asynormal}]
We leverage the same techniques of proving Theorem \ref{th-jointasynormal}.
In other words, we only need to show
\begin{equation} \label{eq-nearmax}
    \ell_n(\ps(\hat{\theta}_n)) \geq \sup_{\theta \in \Theta} \ell_n(\ps(\theta)) - o_{\Prob}(n^{-1}). 
\end{equation}
Note that
\begin{align*}
&\quad\quad \ell_n(\ps(\hat{\theta}_n)) \\
& \geq \ell_n(\hat{p}(\hat{\theta}_n)) - \sup_{\theta \in \Theta} |\ell_n(\hat{p}(\theta)) - \ell_n(\ps(\theta))| \\
& = \sup_{\theta \in \Theta} \ell_n(\hat{p}(\theta))
- \sup_{\theta \in \Theta} |\ell_n(\hat{p}(\theta)) - \ell_n(\ps(\theta))| \\
& \geq \sup_{\theta \in \Theta} [\ell_n(\ps(\theta)) - \sup_{\theta \in \Theta}|\ell_n(\ps(\theta)) - \ell_n(\hat{p}(\theta))|] - \sup_{\theta \in \Theta} |\ell_n(\hat{p}(\theta)) - \ell_n(\ps(\theta))| \\
& \geq \sup_{\theta \in \Theta} \ell_n(\ps(\theta))
- 2 \sup_{\theta \in \Theta} |\ell_n(\hat{p}(\theta)) - \ell_n(\ps(\theta))|. 
\end{align*}
By \eqref{eq-error} and $\omega_n/n^2 \rightarrow \infty$, 
we have
$$
\sup_{\theta \in \Theta}|\ell_n(\hat{p}(\theta)) - \ell_n(\ps(\theta))| = O_{\Prob}\left(\frac{1}{\sqrt{\omega_n}}\right) + o_{\Prob}\left(\frac{1}{n}\right) = o_{\Prob}(n^{-1}). 
$$
Hence, the relation in \eqref{eq-nearmax} is verified. 
By \eqref{eq-equicontinuous}, \eqref{eq-Taylor}, \eqref{eq-nearmax}, and Theorem \ref{th-thetacons}, it follows from Theorem 5.23 in \cite{van2000} that $\sqrt{n}(\hat{\theta}_n - \theta_0)$ is asymptotically normal with mean zero and covariance matrix
$$
V_{\theta_0}^{-1} \left(\frac{\p  \ps(\theta_0)}{\p \theta}\right)^{\prime} \E_{\theta_0} \left[\left\{\frac{\p f(y, \ps(\theta_0))}{\p p}\right\}\left\{\frac{\p f(y, \ps(\theta_0))}{\p p}\right\}^{\prime}\right]\left(\frac{\p  \ps(\theta_0)}{\p \theta}\right) V_{\theta_0}^{-1},
$$
if $V_{\theta_0}$ is non-singular. 
This completes the proof. 
\end{proof}

\section{Proofs in Section \ref{asym-semi}} \label{app-semi}

We first establish an approximation result that helps us to determine an appropriate choice of the sieve space $\mB_n$. 
\begin{lemma} \label{le-appro-semi}
Under Assumptions \ref{ass-compactness-semi} and \ref{ass-ident-semi}, 
we have when $|\tau^{(n)}| = o(1)$,
\begin{align*}
    \sup_{\theta \in \Theta} \inf_{\eta \in \mB_n} \left[\|p^*(\cdot; \theta) - \eta(\cdot)\|_{\infty} \vee \|\Psi(p^*(\cdot; \theta); \theta) - \Psi(\eta(\cdot),\theta)\|_{\infty}\right] = O(|\tau^{(n)}|^4),
\end{align*}
where $a \vee b = \max(a, b)$ denotes the maximum of two real numbers $a$ and $b$.
\end{lemma}

\begin{proof} [Proof of Lemma \ref{le-appro-semi}]
Under Assumption \ref{ass-ident}, $\Psi$ has continuous fourth order derivatives. Therefore, $\frac{\diff^4 p^*(x; \theta)}{\diff x^4}$ is a continuous function of $x$ and $\theta$ for $(x, \theta) \in [0, T] \times \Theta$. Because $\Theta$ is compact, it follows that
$$
\sup_{\theta \in \Theta} \left\| \frac{\diff ^4 p^*(x; \theta)}{\diff x^4} \right\|_{\infty} < \infty.  
$$

Letting $|\tau^{(n)}| \rightarrow 0$ as $n \rightarrow \infty$, by Theorem 2 and Theorem 4 in \cite{hall1976}, we have
\begin{align} \label{eq-psi}
\begin{split}
& \sup_{\theta \in \Theta} \inf_{\eta_{\theta} \in \mB_n} \|p(\cdot; \theta) - \eta_{\theta}\|_{\infty} \leq C_0 \sup_{\theta \in \Theta}  \left\| \frac{\diff ^4 p(x; \theta) }{\diff x^4}\right\|_{\infty} |\tau^{(n)}|^4 \rightarrow 0,\\
& \sup_{\theta \in \Theta} \inf_{\eta_{\theta} \in \mB_n} \|\Psi(p^*(\cdot; \theta); \theta) - \Psi(\eta_{\theta}(\cdot),\theta)\|_{\infty} \\
& \qquad  \leq \sup_{|p| \leq r, \theta \in \Theta} \left| \frac{\diff \Psi}{\diff p}(p; \theta) \right| \times \sup_{\theta \in \Theta} \inf_{\eta_{\theta} \in \mB_n} \|p^*(\cdot; \theta) - \eta_{\theta}\|_{\infty},
\end{split}
\end{align}
for some positive constant $C_0$ and $r$. On the right-hand side of \eqref{eq-psi}, since we choose $\eta_{\theta}$ that best approximates $p^*(x; \theta)$ over $\mB_n$ and $\Theta$ is compact, we can choose a sufficient large but finite $r$ such that we restrict our attention to $p$ which is bounded by $r$ from above. 
Thus, the left-hand side of the bottom line of \eqref{eq-psi} is of order $|\tau^{(n)}|^4$ as well. The proof is completed. 
\end{proof}

\begin{proof} [Proof of Theorem \ref{th-pcons-semi}]
Let $J(p, \theta) = \int_0^T [p(x) - \Psi(p(x),\theta)]^2 \diff x$.
Based on the definition of $\hat{p}_{n}(\cdot, \theta)$, we have  for any $\theta \in \Theta$, 
\begin{align*}
    \ell_n(\hat{p}_{n}(\cdot; \theta)) - \omega_n J(\hat{p}_{n}(\cdot; \theta), \theta) \geq \ell_n(p_{\theta, n}(\cdot)) - \omega_n J(p_{\theta, n}(\cdot), \theta).
\end{align*}
As $\ell_n(\hat{p}_{n}(\cdot; \theta)) \leq 0$ by Assumption \ref{ass-strong-semi}, it follows that
$$
- \omega_n J(\hat{p}_{n}(\cdot; \theta), \theta) \geq \ell_n(p_{\theta, n}(\cdot)) - \omega_n J(p_{\theta, n}(\cdot), \theta).
$$
Then 
\begin{align*}
    &  J(\hat{p}_{n}(\cdot; \theta), \theta) \\
    & \leq -\omega_n^{-1} \ell_n(p_{\theta, n}(\cdot)) +J(p_{\theta, n}(\cdot), \theta)  \\
    & = \omega_n^{-1} \Prob_n[-f(Y, p_{\theta, n}(X))] + \|p_{\theta, n}(\cdot) -\Psi(p_{\theta, n}(\cdot), \theta)\|_{L^2[0,T]}^2\\
    & = \omega_n^{-1} \Prob_n[-f(Y, p_{\theta, n}(X))] + \|p_{\theta, n}(\cdot) -p^*(\cdot; \theta) + \Psi(p^*(\cdot; \theta), \theta) -\Psi(p_{\theta, n}(\cdot), \theta)\|_{L^2[0,T]}^2\\
    & \leq \omega_n^{-1} \Prob_n[-f(Y, p_{\theta, n}(X))] + 2 \|p_{\theta, n}(\cdot) -p^*(\cdot; \theta)\|_{L^2[0, T]}^2 \\
    & \qquad + 2 \|\Psi(p^*(\cdot; \theta), \theta) -\Psi(p_{\theta, n}(\cdot), \theta)\|_{L^2[0,T]}^2\\
    & \leq \omega_n^{-1} \Prob_n[-f(Y, p_{\theta, n}(X))] + 2 \|p_{\theta, n}(\cdot) -p^*(\cdot; \theta)\|_{\infty}^2 \\
    & \qquad + 2 \|\Psi(p^*(\cdot, \theta), \theta) -\Psi(p_{\theta, n}(\cdot), \theta)\|_{\infty}^2\\
    & \leq \omega_n^{-1} \Prob_n[-f(Y, p_{\theta, n}(X)) ] + C_0 r_n^2. 
\end{align*}
for some constant $C_0$, where the last inequality holds by Lemma \ref{le-appro-semi}. 

Note that $\sup_{\theta \in \Theta} |p_{\theta, n}(\cdot)|_{\infty} \leq r$ by the definition of the sieve space $\mB_n(r)$. By invoking the same argument for proving equation \eqref{eq-unifbounded}, we have 
$$
\sup_{\theta \in \Theta} \Prob_n[-f(Y, p_{\theta, n}(X)) ] = O_{\Prob}(1).
$$
It follows that
\begin{align*}
  \sup_{\theta \in \Theta} J(\hat{p}_{n}(\cdot; \theta), \theta)
   & \leq \omega_n^{-1} O_{\Prob}(1) + C_0 r_n^2 \\
   & = O_{\Prob}(\omega_n^{-1}) + O(r_n^2).
\end{align*}

As $J(p, \theta) = \int_0^T [p(x) - \Psi(p(x),\theta)]^2 \diff x$, we have 
$$
\sup_{\theta \in \Theta} \|\hat{p}_{n}(\cdot; \theta) - \Psi(\hat{p}_{n}(\cdot; \theta), \theta) \| _{L^2[0, T]}^2 \leq O_{\Prob}(\omega_n^{-1}) + O(r_n^2).
$$
Since $p^*(x; \theta)$ satisfies $p(x) = \Psi(p(x), \theta)$,
$$
\hat{p}_{n}(x; \theta) - \Psi(\hat{p}_{n}(x; \theta), \theta) 
= g(p^*(x; \theta), \theta) - g(\hat{p}_{n}(x; \theta), \theta)
$$
holds for for any $x \in [0, T]$ and $\theta \in \Theta$. Moreover, under Assumption \ref{ass-ident-semi},  we have 
\begin{align*}
\begin{split}
& \sup_{\theta \in \Theta} \quad \|\hat{p}_{n}(\cdot; \theta) - p^*(\cdot; \theta)\| _{L^2[0, T]}^2 \\
& \leq  \sup_{\theta \in \Theta} C_g^2 \|g(p^*(\cdot; \theta), \theta) - g(\hat{p}_{n}(\cdot; \theta), \theta)\| _{L^2[0, T]}^2 \\
& = \sup_{\theta \in \Theta} C_g^2 \|\hat{p}_{n}(\cdot; \theta) - \Psi(\hat{p}_{n}(\cdot; \theta), \theta)\|_{L^2[0, T]}^2 \\
& = O_{\Prob}(\omega_n^{-1}) + O(r_n^2).
\end{split}
\end{align*}
Lastly, under Assumption \ref{ass-density}, one obtains
$$
 \sup_{\theta \in \Theta}  \|\hat{p}_n(\cdot; \theta) - p^*(\cdot; \theta)\|_{L^2(\Prob)}^2 = O_{\Prob}(\omega_n^{-1}) + O(r_n^2).
$$
The proof is completed. 

\end{proof}

To apply Theorem 5.23 of \cite{van2000} to show asymptotic normality for $\hat{\theta}_n$, we need the following lemma. 
\begin{lemma} \label{lem:applk-semi}
Under the same conditions in Theorem \ref{th-nestasynormal-semi}, we have
\begin{equation} \label{eq-nearmaxjoint-semi}
    \ell_n(\ps(\cdot; \hat{\theta}_n)) \geq \sup_{\theta \in \Theta} \ell_n(\ps(\cdot; \theta)) - o_{\Prob}(n^{-1}),
\end{equation}
and 
\begin{equation} \label{eq-uWLLN-semi}
    \sup_{\theta \in \Theta} |\ell_n(\ps(\cdot; \theta)) - M(\theta)| = o_{\Prob}(1). 
\end{equation}
\end{lemma}

\begin{proof}[Proof of Lemma \ref{lem:applk-semi}]
Under Assumption \ref{ass-normal-semi}, there exists a constant $\delta > 0$, such that
\begin{equation} \label{eq-minmax-semi}
\sup_{|p| \leq r + 1} \left|\frac{\p f}{\p p}(y, p) \right| \leq \delta^{-1} \left\{1 + \inf_{|p| \leq r + 1} \left|\frac{\p f}{\p p}(y, p) \right|\right\}.
\end{equation}
Then it follows 
\begin{align} \label{eq-nllkerror-semi}
\begin{split}
& \qquad  |\ell_n(\hat{p}_n(\cdot; \theta)) - \ell_n (\ps(\cdot; \theta))| \\
& \leq \frac{1}{n} \sum_{i = 1}^n |f(Y_i, \hat{p}_n(X_i; \theta)) - f(Y_i, \ps(X_i; \theta)) | \\
& \leq \frac{1}{n} \sum_{i = 1}^n \left\{ \sup_{|p| \leq r + 1} \left|\frac{\p f}{\p p}(Y_i, p) \right|\right\} |\hat{p}_n(X_i; \theta) - \ps(X_i; \theta) | \\
& \leq \frac{1}{n} \sum_{i = 1}^n \frac{1}{\delta} \left\{ 1 + \inf_{|p| \leq r + 1} \left|\frac{\p f}{\p p}(Y_i, p) \right|\right\} |\hat{p}_n(X_i; \theta) - \ps(X_i; \theta) | \\
& \leq \frac{1}{n} \sum_{i = 1}^n \left\{ 1 + \left|\frac{\p f}{\p p}(Y_i, \ps(X_i; \theta_0)) \right|\right\} |\hat{p}_n(X_i; \theta) - \ps(X_i; \theta) |.
\end{split}
\end{align}

Next we identify the stochastic order of the right-hand side of \eqref{eq-nllkerror-semi} using empirical process theory. To simplify the notation, for any function $\eta \in \mB_n(r)$ and $\theta \in \Theta$, we denote 
$H(r, \eta, \theta)$ as the right-hand side function from one single observation with parameter $(\eta, \theta)$, i.e., 
$$
H(r, \eta, \theta) = \left\{ 1 + \left|\frac{\p f}{\p p}(y, \ps(x; \theta_0)) \right|\right\} |\eta(x) - \ps(x; \theta) |. 
$$
Then we define $$
\mathbb{G}_n [H(R, \hat{p}_n(\cdot; \theta), \theta)] = \sqrt{n}(\Prob_n - \Prob) [H(R, \hat{p}_n(\cdot; \theta), \theta)]. 
$$
To find the upper bound on $\mathbb{G}_n [H(R, \hat{p}_n(\cdot; \theta), \theta)] $, 
we consider a function class $\ca L_n$ defined by $\{H(r, \eta, \theta): \eta \in \mB_n(r), \theta \in \Theta\}$. By Assumption \ref{ass-normal-semi} and the definition of the sieve space $\mB_n(r)$, the class $\ca L_n$ has an upper bound $O_{\Prob}(\log n)$. Additionally, this class can be treated as a class of functions indexed by $\theta$ and $\{\beta_j\}_{j = 1}^{K_n}$, which are the B-spline coefficients of $\eta$ in $\mB_n(r)$. It is straightforward to verify that  functions in $\ca L_n$ are Lipschitz continuous with respect to all parameters and the Lipschitz constant is bounded by $O_{\Prob}(\log n)$. Additionally, since $\theta$ is bounded by some constant, and all these B-spline coefficients are bounded by $r$, they must reside in a hypercube of $\Re^{K_n + 1}$. Hence, if we partition this large hypercube into a set of smaller hypercubes with scale length $\epsilon$, the cardinality number of this set is no more than $O(\epsilon^{-K_n})$. Furthermore, by the Lipschitz property of functions in $\ca L_n$, the $L_{\infty}$ distance between any two functions in the same subcube is bounded by $O_{\Prob}(\log n) K_n \epsilon$. Therefore, the bracketing number \citep[cf.][p. 270]{van2000} of $\ca L_n$ satisfies $N_{[\cdot]} (O_{\Prob}(\log n) K_n \epsilon, \ca L_n, L_{\infty}) \leq O(1) \epsilon^{-K_n}$. 
Then by Theorem 19.35 of \cite{van2000} and $n^{1/4} = o(K_n)$, we have in probability
\begin{align*}
\sqrt{n} \mathbb{E}^*\|\Prob_n - \Prob\|_{\ca L_n} & \leq O_{\Prob}(1) \int_{0}^1 \sqrt{\log \left\{\frac{2 (\log n) K_n}{\epsilon}\right\}^{2K_n}} \diff \epsilon \\
& \leq O_{\Prob}(1) K_n^{1/2} \sqrt{\log(K_n)}. 
\end{align*}
Therefore, $\mathbb{G}_n [H(R, \hat{p}_n(\cdot; \theta), \theta)]$  is bounded by $O_{\Prob}(K_n^{1/2} \sqrt{\log(K_n)} /\sqrt{n})$ from above, which is $o_{\Prob}(1)$ by the assumption $K_n^2 \log (K_n) = o(n)$. 

Furthermore, with some abuse of notation, we still use $\ca L_n$ to denote the class 
$$
\left\{H(r, \theta): H(r, \theta) =  \left\{ 1 + \left|\frac{\p f}{\p p}(y, \ps(x; \theta_0)) \right|\right\} |\hat{p}_n(x; \theta) - \ps(x; \theta) | \right\}. 
$$
Obviously, the index set $\Theta$ is totally bounded when equipped with the Euclidean distance. Moreover, we can choose $F_n = (2r + 1) \{1 + |{\p f}(y, \ps(x; \theta_0)/{\p p}\}$ to be the envelop function of $\ca L_n$, which has finite moments of all orders. Additionally, from the preceding arguments, we know this class is equi-continuous with respect to $\theta$. Therefore, by Theorem 2.11.23 of \cite{van1996}, $\mathbb{G}_n [H(R, \hat{p}_n(\cdot; \theta), \theta)]$ is bounded by $o_{\Prob}(1/n)$ from above. Consequently, it follows from Theorem \ref{th-pcons-semi} that
\begin{equation} \label{eq-uniform-llkapproxmiation-semi}
\sup_{\theta \in \Theta}|\ell_n(\hat{p}_n(\cdot; \theta)) - \ell_n (\ps(\cdot; \theta))|   = o_{\Prob}(n^{-1})
\end{equation}
if $r_n = o(n^{-1})$ and $\omega_n /n^2 \rightarrow \infty$.

By definition of $\ps(\cdot; \theta)$ and $\hat{\theta}_n$, we have
\begin{align*}
&\quad\quad \ell_n(\ps(\cdot; \hat{\theta}_n)) \\
& \geq \ell_n(\hat{p}_n(\cdot; \hat{\theta}_n)) - \sup_{\theta \in \Theta} |\ell_n(\hat{p}_n(\cdot; \theta)) - \ell_n(\ps(\cdot; \theta))| \\
& \geq \sup_{\theta \in \Theta} \ell_n(\ps(\cdot; \theta))
- 2\sup_{\theta \in \Theta} |\ell_n(\hat{p}_n(\cdot; \theta)) - \ell_n(\ps(\cdot; \theta))|. 
\end{align*}
Hence, the relation in \eqref{eq-nearmaxjoint-semi} holds. Equation \eqref{eq-uWLLN-semi} can be obtained with a slight modification of Lemma \ref{le-unifSLLN}. 
\end{proof}

Next we prove Theorem \ref{th-consistency-semi}.
\begin{proof} [Proof of Theorem \ref{th-consistency-semi}]
We employ the techniques of proving Theorem \ref{th-jointcons}.
By \eqref{eq-uWLLN-semi} and \eqref{eq-uniform-llkapproxmiation-semi}, we have
  \begin{align*}
    & \quad\quad M(\theta_0) - M(\hat{\theta}_n) \\
    & = M_n(\theta_0) - M_n(\hat{\theta}_n) + o_{\Prob}(1)\quad\text{by}~\eqref{eq-uWLLN-semi} \\
    & = \ell_n(\ps(\cdot; \theta_0)) - \ell_n(\ps(\cdot; \hat{\theta}_n)) + o_{\Prob}(1)\\
    & \leq \ell_n(\hat{p}_n(\cdot; \theta_0)) + \sup_{\theta \in \Theta}|\ell_n(\hat{p}(\cdot; \theta)) - \ell_n(\ps(\cdot; \theta))| - \ell_n(\hat{p}(\cdot; \hat{\theta}_n)) \\
    & \quad \quad + \sup_{\theta \in \Theta}|\ell_n(\hat{p}(\cdot; \theta)) - \ell_n(\ps(\cdot; \theta))| + o_{\Prob}(1) \\
    & = \ell_n(\hat{p}(\cdot; \theta_0)) - \ell_n(\hat{p}(\cdot; \hat{\theta}_n)) 
    + 2\sup_{\theta \in \Theta}|\ell_n(\hat{p}(\cdot; \theta)) - \ell_n(\ps(\cdot; \theta))| + o_{\Prob}(1) \\
    & \leq \ell_n(\hat{p}(\cdot; \theta_0)) - \ell_n(\hat{p}(\cdot; \hat{\theta}_n)) + o_{\Prob}(1).  \quad\text{by}~\eqref{eq-uniform-llkapproxmiation-semi}
\end{align*}
By definition of $\hat{\theta}_n$, we have
$$
\ell_n(\hat{p}(\cdot; \hat{\theta}_n)) \geq \ell_n(\hat{p}(\cdot; \theta_0)),
$$
so $M(\theta_0) - M(\hat{\theta}_n) \leq o_{\Prob}(1)$. By \eqref{eq-Mest}, we have
$$
\{d(\hat{\theta}_n, \theta_0) \geq \delta\} \subset \{M(\theta_0) - M(\hat{\theta}_n) \geq \gamma\} \subset \{o_{\Prob}(1) \geq \gamma\}.
$$
Therefore, 
$$
\Prob_{\theta_0}(d(\hat{\theta}_n, \theta_0) \geq \delta) \leq \Prob_{\theta_0} (o_{\Prob}(1) \geq \gamma),
$$
which converges to 0 as $n$ approaches infinity. As $\delta$ is an arbitrary positive number, $\hat{\theta}_n$ is a consistent estimator of $\theta_0$. This completes the proof. \end{proof}

Now we are ready to prove Theorem \ref{th-nestasynormal-semi}.
\begin{proof}[Proof of Theorem \ref{th-nestasynormal-semi}]
Similar to the proof of Theorem \ref{th-jointasynormal}, 
we follow Theorem 5.23 of \cite{van2000} to prove asymptotic normality of $\hat{\theta}_n$.
Firstly, as we have shown in the proof of Lemma \ref{le-unifSLLN},  under Assumption \ref{ass-normal-semi}, we have
\begin{align}
    \nonumber
    &\quad\quad |f(y, \ps(x; \theta_1)) - f(y, \ps(x; \theta_2))| \\
    \nonumber
    & \leq \frac{1}{\eta} \left[1 +  \left|\frac{\p f}{\p p}(y, \ps(x; \theta_0)) \right|\right] |\ps(\theta_1) - \ps(\theta_2)| \\
    & \leq \frac{C}{\eta} \left[1 +  \left|\frac{\p f}{\p p}(y, \ps(x; \theta_0)) \right|\right] \|\theta_1 - \theta_2\|_2
    \label{eq-equicontinuous-semi}
\end{align}
for some constant $C$, and $\eta$ is defined in Equation \eqref{eq-eta}. By Assumption \ref{ass-normal-semi}, the right-hand side of \eqref{eq-equicontinuous-semi} has a finite second moment.

Next, we consider a second-order Taylor expansion for
$$
M(\theta) = \E_{\theta_0}[f(Y, \ps(X; \theta))]
$$
in a neighbourhood of $\theta_0$. Obviously, 
\begin{align} \label{eq-linterms-semi}
\begin{split}
&\quad\quad f(y, p(x; \theta)) \\
& = f(y, \ps(x; \theta_0)) + \left[\frac{\p f(y, \ps(x; \theta_0))}{\p p}\right]^{\prime}\left(\frac{\p  \ps(x; \theta_0)}{\p \theta}\right)^{\prime} (\theta - \theta_0) \\
&\quad + \frac{1}{2} (\theta-\theta_0)^{\prime} \left[\frac{\p f(y, \ps(x; \theta_0))}{\p p}
\frac{\p^2  \ps(x; \theta_0)}{\p \theta \p \theta^{\prime}} \right.\\
& \qquad \left. + \frac{\p^2 f(y, \ps(x; \theta_0))}{\p p^2} \left(\frac{\p  \ps(x; \theta_0)}{\p \theta}\right) \left(\frac{\p  \ps(x; \theta_0)}{\p \theta}\right) ^{\prime}\right]  \\
& \quad \times (\theta - \theta_0) + R,
\end{split}
\end{align}
where $R$ is the remainder term. Define
\begin{align*}
\begin{split}
D(y, x, \theta) & = \frac{\p f(y, \ps(x; \theta_0))}{\p p}
\frac{\p^2  \ps(x; \theta_0)}{\p \theta \p \theta^{\prime}} \\
& \qquad + \frac{\p^2 f(y, \ps(x; \theta_0))}{\p p^2} \left(\frac{\p  \ps(x; \theta_0)}{\p \theta}\right) \left(\frac{\p  \ps(x; \theta_0)}{\p \theta}\right) ^{\prime}. 
\end{split}
\end{align*}
Then the reminder term can be rewritten as
$$
R = (\theta - \theta_0)^{\prime} \left[\int_0^1 [D(y, x, \theta_0 + s(\theta - \theta_0)) - D(y, x, \theta_0)] (1 - s) \diff s \right] (\theta - \theta_0).
$$
Note that $D(y, x, \theta)$ is a $d \times d$ matrix. For any $(a,b)$th entry in $D(y, x, \theta)$, by using the same argument for Equation \eqref{eq-maxbound},
we can show that, for any $\theta \in \Theta$,
\begin{align*}
 & \quad\quad  |D_{ab}(y, x, \theta)| \\
 & \leq \left|\frac{\p f(y, \ps(x; \theta))}{\p p}
\frac{\p^2  \ps(x; \theta)}{\p \theta_a \p \theta_b}\right| + \left|\frac{\p^2 f(y, \ps(x; \theta))}{\p p^2} \left(\frac{\p  \ps(x; \theta)}{\p \theta_a}\right)^{\prime} \left(\frac{\p  \ps(x; \theta)}{\p \theta_b}\right)\right| \\
& \leq \frac{C^{\prime}}{\eta} \left[1 +  \left|\frac{\p f}{\p p}(y, \ps(x; \theta_0)) \right| + \left|\frac{\p^2 f}{\p p^2}(y, \ps(x; \theta_0))\right|\right],
\end{align*}
where $C^\prime$ is a positive constant. Additionally, under Assumption \ref{ass-normal-semi}, the right-hand side of the above inequality has a finite mean. Therefore, applying the dominated convergence theorem, we have
$$
\E_{\theta_0} \left[\int_0^1 [D(y, x, \theta_0 + s(\theta - \theta_0)) - D(y, x,\theta_0)] (1 - s) \diff s \right] \rightarrow 0
$$
as $\theta \rightarrow \theta_0$. Then, by the Taylor expansion of $f(y, \ps(x; \theta))$, it follows that 
\begin{equation} \label{eq-Taylor-semi}
    M(\theta) = M(\theta_0) + \frac{1}{2} (\theta - \theta_0)^{\prime} V_{\theta_0} (\theta - \theta_0) + o\left(\|\theta - \theta_0\|_2^2\right).
\end{equation}
Recall that $f(y, p)$ is the log density of $Y_i$.
Therefore, there is no linear form in \eqref{eq-Taylor-semi} and the expected value of $D(Y, X, \theta_0)$ is given by $V_{\theta_0}$
as the expectation of \eqref{eq-linterm} is zero.

Finally, \eqref{eq-nearmaxjoint-semi} holds by Lemma \ref{lem:applk-semi}. Moreover, we can easily show $\hat{\theta}_n$ is a consistent estimator of $\theta_0$ from \eqref{eq-uWLLN-semi} in Lemma \ref{lem:applk-semi}.
Combining \eqref{eq-equicontinuous-semi}, \eqref{eq-Taylor-semi} and \eqref{eq-nearmaxjoint-semi}, it follows from Theorem 5.23 in \cite{van2000} that $\sqrt{n}(\hat{\theta}_n - \theta_0)$ is asymptotically normal with mean zero and covariance matrix
$$
V_{\theta_0}^{-1}  \E_{\theta_0} \left[\left\{\frac{\p f(Y, \ps(X; \theta_0))}{\p p}\right\}^2 \left(\frac{\p  \ps(X; \theta_0)}{\p \theta}\right)^{\prime} \left(\frac{\p  \ps(X; \theta_0)}{\p \theta}\right) \right] V_{\theta_0}^{-1},
$$
if $V_{\theta_0}$ is non-singular. 
This completes the proof. 
\end{proof}

\section{Additional Simulations} \label{addsimu}

The DGP is only for demonstration purposes in the Monte Carlo simulations in the main text. We now conduct additional simulation experiments in a rich setting building on our empirical application to further demonstrate our method. Specifically, we constructed a richer DGP using market- and player-specific variables along with the maximum likelihood estimates. We set the true parameters to the rounded estimates. We sampled markets (with replacement) using the same dataset and re-generated Walmart and Kmart's choices using the equilibrium CCPs. We experimented with different sample sizes, $N=2000$ and $N=4000$.

\begin{landscape}
\begin{table}[h]
    \centering
    \caption{Simulation Results}
    \begin{adjustbox}{width=\linewidth,center}
    \begin{threeparttable}
\begin{tabular}{l r | rrrr | rrrr} \\ \hline \hline
   & & \multicolumn{2}{c}{N=2000} & \multicolumn{2}{c}{N=4000} & \multicolumn{2}{c}{N=2000} & \multicolumn{2}{c}{N=4000} \\ \hline
             & $\theta$ & Ours & std   & Ours & std   & MLE & std   & MLE & std   \\ \hline   
               \textbf{Market-specific} &&&& & \\  
pop & 3   & 3.0274   & 0.1280 & 2.9987   & 0.0968 & 3.0309   & 0.1301 & 3.0018   & 0.0933 \\
spc          & 3   & 3.0117   & 0.1924 & 3.0092   & 0.1479 & 3.0217   & 0.1913 & 3.0153   & 0.1448 \\
urban        & 2   & 2.0218   & 0.2914 & 1.9918   & 0.2130 & 2.0262   & 0.2962 & 1.9953   & 0.2104 \\
\textbf{Walmart-specific} &&&& & \\
intercept  & -22 & -22.2541 & 1.7524 & -22.1481 & 1.4207 & -22.3295 & 1.7184 & -22.1859 & 1.3371 \\
dbenton  & -2  & -1.9875  & 0.1461 & -1.9854  & 0.1039 & -1.9907  & 0.1407 & -1.9896  & 0.0974 \\
south     & 1   & 1.0008   & 0.1560 & 0.9820   & 0.1100 & 1.0067   & 0.1473 & 0.9868   & 0.1018 \\
\textbf{Kmart-specific} &&&& & \\
intercept\_K & -36 & -36.1569 & 1.6444 & -36.0579 & 1.3321 & -36.2479 & 1.6475 & -36.1200 & 1.3150 \\
midwest    & 1   & 1.0073   & 0.1452 & 0.9992   & 0.0989 & 1.0006   & 0.1437 & 0.9957   & 0.0965 \\ \hline 
$\Delta$      & 2   & 2.0605   & 0.2954 & 2.0178   & 0.2337 & 2.0620   & 0.2441 & 2.0133   & 0.1802 \\ \hline
\end{tabular}
    \begin{tablenotes}
\item Note: We estimate the model using the proposed method when $K=10,20,30$ and obtain very similar results. We report only the results for $K=10$ in this table. 
\end{tablenotes}
    \end{threeparttable}
    \end{adjustbox}
\end{table}
\end{landscape}

\section{Deriving Gradient and Hessian Functions}

\subsection{The Monopoly Pricing Problem} \label{JHprice}

In this subsection, we derive the gradient of the objective function in the monopoly pricing problem. 

In the inner loop, we maximize the following function with respect to $\beta$
\[
h(\beta,\theta) = 
\sum \log \phi(y_i - p^\beta(x_i))
- \omega \sum_{\ell=1}^{L} \left[p^\beta(x_\ell) e^{p^\beta(x_\ell)} - \theta x_\ell\right]^2,
\]
where $L=1000$ is the number of grid points to approximate the integration. Note that the standard normal density $\phi(z) = \frac{1}{\sqrt{2\pi}} \exp{(-\frac{z^2}{2})}$ and $p^\beta(x) = \sum_k \beta_k s_k(x)$. We will consider the two terms in sequence.

In the first step, 
 \begin{align*}
    \frac{\partial \ell}{\partial \beta_k} =& \sum (y_i - p^\beta(x_i)) s_k(x_i) ,\\
    \frac{\partial^2 \ell}{\partial \beta_k \partial \beta_{k^\prime}} =& - \sum s_k(x_i) s_{k^\prime}(x_i).
\end{align*}

In the second step, 
 \begin{align*}
 \begin{split}
    \frac{\partial \rho}{\partial \beta_k} =& 2 \sum_{\ell=1}^{L} [p^\beta(x_\ell) e^{p^\beta(x_\ell)} - \theta x_\ell]e^{p^\beta(x_\ell)} [1+p^\beta(x_\ell)]s_k(x_\ell), \\
    \frac{\partial^2 \rho}{\partial \beta_k \partial \beta_{k^\prime}} =& 2 \sum_{\ell=1}^{L} \Big[\big(e^{p^\beta(x_\ell)} [1+p^\beta(x_\ell)] \big)^2 \\
    & \qquad +[p^\beta(x_\ell) e^{p^\beta(x_\ell)} - \theta x_\ell]e^{p^\beta(x_\ell)} [2+p^\beta(x_\ell)]
    \Big] s_k(x_\ell) s_{k^\prime}(x_\ell) ,\\
     \frac{\partial^2 \rho}{\partial \beta_k \partial \theta} =& 
     -2 \sum_{\ell=1}^{L} x_\ell e^{p^\beta(x_\ell)} [1+p^\beta(x_\ell)]s_k(x_\ell) .
\end{split}
\end{align*}

\subsection{Static Game with Incomplete Information} \label{JHjia}

In this subsection, we derive the gradient of the objective function in the static game with incomplete information, where  
\begin{align}
\ell(\beta, \theta) & = 
\sum_{j=\W,\K}\sum_{m=1}^{M}\left[d_{jm}\log\left(p_{j}^\beta(\xi_{m}(\theta))\right)+(1-d_{jm})\log\left(1-p_{j}^\beta(\xi_{m}(\theta))\right)\right] \\
\rho(\beta, \theta)  & = \sum_{j=\W,\K}\sum_{m=1}^{M} \left[p_j^\beta(\xi_{m}(\theta)) - \sigma \Big( \xi_{jm} - p_{-j}^\beta(\xi_{m}(\theta)) \Delta \Big) \right]^2    
\end{align}
We approximate the CCP by $p_j^\beta (\xi_{j},\xi_{-j}) = \sigma \Big(\sum_{\imath=1}^K\sum_{\jmath=1}^K\beta_{\imath \jmath} s_\imath \big(\sigma(\xi_{j}) \big) s_{\jmath} \big(\sigma(\xi_{-j}) \big) \Big)$. Denote $v_j = \sum_{\imath=1}^K\sum_{\jmath=1}^K\beta_{\imath \jmath} s_\imath \big(\sigma(\xi_{j}) \big) s_{\jmath} \big(\sigma(\xi_{-j}) \big)$ for convenience. Note that $\sigma(v) = \frac{1}{1+e^{-v}}$ and $\sigma^\prime(v) = \frac{-1}{\sigma^2} e^{-v} (-1) = \frac{e^{-v}}{\sigma(v)^2}$. Moreover, $\xi_{jm} = X_{jm}^\prime \alpha - Z_m^\prime \gamma$. We stack the parameters $\theta = (\gamma,\alpha_w,\alpha_k,\Delta)^\prime$, where $\alpha_w$ and $\alpha_k$ are WalMart and Kmart's coefficients.  

First, denote $p_{jm} = p_j^\beta(\xi_m(\theta))$. 
\begin{align*}
\frac{\partial \ell}{\partial \beta_{\imath \jmath}} = & 
\sum_{j=\W,\K}\sum_{m=1}^{M} \left[ \frac{d_{jm}}{p_{jm}}- \frac{1-d_{jm}}{1-p_{jm}}\right] \cdot
\frac{\partial p_{jm}}{\partial \beta_{\imath \jmath}} , \\
\frac{\partial \ell}{\partial \theta_k} = & 
\sum_{j=\W,\K}\sum_{m=1}^{M} \left[ \frac{d_{jm}}{p_{jm}}- \frac{1-d_{jm}}{1-p_{jm}}\right] \cdot
\frac{\partial p_{jm}}{\partial \theta_k} ,  \\
\frac{\partial \ell}{\partial \Delta} = & 0 ,  \\
\frac{\partial \rho}{\partial \beta_{\imath \jmath}}
=& 2\sum_{j=\W,\K}  \sum_{m=1}^{M} \Big[p_{jm} - \sigma(\eta_{jm}) \Big] \Big[\frac{\partial p_{jm}}{\partial \beta_{\imath \jmath}}  + \Delta \sigma^\prime(\eta_{jm}) \frac{\partial p_{-jm}}{\partial \beta_{\imath \jmath}} \Big], \\
\frac{\partial \rho}{\partial \theta_k}
=& 2\sum_{j=\W,\K}  \sum_{m=1}^{M} \Big[p_{jm} - \sigma(\eta_{jm}) \Big] \Big[\frac{\partial p_{jm}}{\partial \theta_k}  - \sigma^\prime(\eta_{jm}) (\frac{\partial \xi_{jm}}{\partial \theta_k} - \Delta  \frac{\partial p_{-jm}}{\partial \theta_k}) \Big], \\
\frac{\partial \rho}{\partial \Delta}
=& 2\sum_{j=\W,\K}  \sum_{m=1}^{M} \Big[p_{jm} - \sigma(\eta_{jm}) \Big] \Big[ \sigma^\prime(\eta_{jm}) p_{-jm} \Big], 
\end{align*}
where $\eta_{jm}=\xi_{jm} - p_{-j}^\beta(\xi_{m}(\theta)) \Delta$, 
\begin{align*}
    \frac{\partial p_{jm}}{\partial \beta_{\imath \jmath}} = & \sigma^\prime(v_{jm}) s_\imath \big(\sigma(\xi_{jm}) \big) s_{\jmath} \big(\sigma(\xi_{-jm}) \big) \\
    \frac{\partial p_{jm}}{\partial \theta_k} = & \sigma^\prime(v_{jm}) \Big[\sum_{\imath=1}^K\sum_{\jmath=1}^K\beta_{\imath \jmath} \big(
s_\imath^\prime s_{\jmath} \sigma^\prime(\xi_{jm}) \frac{\partial \xi_{jm}}{\partial \theta_k} + 
s_\imath s_{\jmath}^\prime \sigma^\prime(\xi_{-jm}) \frac{\partial \xi_{-jm}}{\partial \theta_k} \big)
\big) \Big].
\end{align*}

Note that $\frac{\partial \ell}{\partial \Delta} = 0$, which implies that the data likelihood is independent of $\Delta$. To make the data likelihood depend on $\theta$ more explicitly, we could replace $p_j^\beta$ by $\sigma(\xi_{jm}-p_{-j}^\beta(\xi_{m}(\theta) \Delta))$. 
\begin{align}
    \frac{\partial \ell}{\partial \theta_k} = & 
\sum_{j=\W,\K}\sum_{m=1}^{M} \left[ \frac{d_{jm}}{\sigma_{jm}}- \frac{1-d_{jm}}{1-\sigma_{jm}}\right] \cdot
\frac{\partial \sigma_{jm}}{\partial \theta_k} , \\
\frac{\partial \sigma_{jm}}{\partial \theta_k} = & 
\sigma^\prime(\eta_{jm}) \Big[\frac{\partial \xi_{jm}}{\partial \theta_k} - \frac{\partial p_{-jm}}{\partial \theta_k} \Delta \Big] \\
\frac{\partial \sigma_{jm}}{\partial \Delta} = & 
\sigma^\prime(\eta_{jm}) \Big[ - p_{-j}^\beta(\xi_{m}(\theta) \Big] 
\end{align}
where $\sigma_{jm} = \sigma(\xi_{jm}-p_{-j}^\beta(\xi_{m}(\theta) \Delta))$ represents the best response to the opponent $-j$.  

\newpage 

\singlespacing

\bibliographystyle{ecta}
\bibliography{refHLE}

\begin{thebibliography}{33}
\newcommand{\enquote}[1]{``#1''}
\expandafter\ifx\csname natexlab\endcsname\relax\def\natexlab#1{#1}\fi

\bibitem[\protect\citeauthoryear{Aguirregabiria and Mira}{Aguirregabiria and
  Mira}{2002}]{aguirregabiria2002swapping}
\textsc{Aguirregabiria, V. and P.~Mira} (2002): \enquote{Swapping the nested
  fixed point algorithm: A class of estimators for discrete Markov decision
  models,} \emph{Econometrica}, 70, 1519--1543.

\bibitem[\protect\citeauthoryear{Aguirregabiria and Mira}{Aguirregabiria and
  Mira}{2007}]{aguirregabiria2007sequential}
---\hspace{-.1pt}---\hspace{-.1pt}--- (2007): \enquote{Sequential estimation of
  dynamic discrete games,} \emph{Econometrica}, 75, 1--53.

\bibitem[\protect\citeauthoryear{Bajari, Hong, Krainer, and Nekipelov}{Bajari
  et~al.}{2010}]{bajari2010estimating}
\textsc{Bajari, P., H.~Hong, J.~Krainer, and D.~Nekipelov} (2010):
  \enquote{Estimating static models of strategic interactions,} \emph{Journal
  of Business \& Economic Statistics}, 28, 469--482.

\bibitem[\protect\citeauthoryear{Barwick and Pathak}{Barwick and
  Pathak}{2015}]{barwick2015costs}
\textsc{Barwick, P.~J. and P.~A. Pathak} (2015): \enquote{The costs of free
  entry: an empirical study of real estate agents in Greater Boston,} \emph{The
  RAND Journal of Economics}, 46, 103--145.

\bibitem[\protect\citeauthoryear{Berry, Levinsohn, and Pakes}{Berry
  et~al.}{1995}]{berry1995automobile}
\textsc{Berry, S., J.~Levinsohn, and A.~Pakes} (1995): \enquote{Automobile
  prices in market equilibrium,} \emph{Econometrica}, 841--890.

\bibitem[\protect\citeauthoryear{Bresnahan and Reiss}{Bresnahan and
  Reiss}{1991}]{bresnahan1991empirical}
\textsc{Bresnahan, T.~F. and P.~C. Reiss} (1991): \enquote{Empirical models of
  discrete games,} \emph{Journal of Econometrics}, 48, 57--81.

\bibitem[\protect\citeauthoryear{Chen, Chen, and Tamer}{Chen
  et~al.}{2023{\natexlab{a}}}]{chen2021efficient}
\textsc{Chen, J., X.~Chen, and E.~Tamer} (2023{\natexlab{a}}):
  \enquote{Efficient estimation in {NPIV} models: {Simulation} comparisons of
  neural network estimators,} \emph{Journal of Econometrics}, 235, 1848--1875.

\bibitem[\protect\citeauthoryear{Chen}{Chen}{2007}]{chen2007large}
\textsc{Chen, X.} (2007): \enquote{Large sample sieve estimation of
  semi-nonparametric models,} \emph{Handbook of Econometrics}, 6, 5549--5632.

\bibitem[\protect\citeauthoryear{Chen, Gentry, Li, and Lu}{Chen
  et~al.}{2023{\natexlab{b}}}]{chen2023identification}
\textsc{Chen, X., M.~L. Gentry, T.~Li, and J.~Lu} (2023{\natexlab{b}}):
  \enquote{Identification and inference in first-price auctions with risk
  averse bidders and selective entry,} \emph{Available at SSRN 3681530}.

\bibitem[\protect\citeauthoryear{Dub{\'e}, Fox, and Su}{Dub{\'e}
  et~al.}{2012}]{dube2012improving}
\textsc{Dub{\'e}, J.-P., J.~T. Fox, and C.-L. Su} (2012): \enquote{Improving
  the numerical performance of static and dynamic aggregate discrete choice
  random coefficients demand estimation,} \emph{Econometrica}, 80, 2231--2267.

\bibitem[\protect\citeauthoryear{Guerre, Perrigne, and Vuong}{Guerre
  et~al.}{2000}]{guerre2000optimal}
\textsc{Guerre, E., I.~Perrigne, and Q.~Vuong} (2000): \enquote{Optimal
  nonparametric estimation of first-price auctions,} \emph{Econometrica}, 68,
  525--574.

\bibitem[\protect\citeauthoryear{Hall and Meyer}{Hall and
  Meyer}{1976}]{hall1976}
\textsc{Hall, C.~A. and W.~W. Meyer} (1976): \enquote{Optimal error bounds for
  cubic spline interpolation,} \emph{Journal of Approximation Theory}, 16,
  105--122.

\bibitem[\protect\citeauthoryear{Hotz and Miller}{Hotz and
  Miller}{1993}]{hotz1993conditional}
\textsc{Hotz, V.~J. and R.~A. Miller} (1993): \enquote{Conditional choice
  probabilities and the estimation of dynamic models,} \emph{The Review of
  Economic Studies}, 60, 497--529.

\bibitem[\protect\citeauthoryear{Iskhakov, Lee, Rust, Schjerning, and
  Seo}{Iskhakov et~al.}{2016}]{iskhakov2016comment}
\textsc{Iskhakov, F., J.~Lee, J.~Rust, B.~Schjerning, and K.~Seo} (2016):
  \enquote{Comment on “constrained optimization approaches to estimation of
  structural models”,} \emph{Econometrica}, 84, 365--370.

\bibitem[\protect\citeauthoryear{Jia}{Jia}{2008}]{jia2008happens}
\textsc{Jia, P.} (2008): \enquote{What happens when Wal-Mart comes to town: An
  empirical analysis of the discount retailing industry,} \emph{Econometrica},
  76, 1263--1316.

\bibitem[\protect\citeauthoryear{Keane and Wolpin}{Keane and
  Wolpin}{1994}]{keane1994solution}
\textsc{Keane, M.~P. and K.~I. Wolpin} (1994): \enquote{The solution and
  estimation of discrete choice dynamic programming models by simulation and
  interpolation: Monte Carlo evidence,} \emph{The Review of Economics and
  Statistics}, 648--672.

\bibitem[\protect\citeauthoryear{Keane and Wolpin}{Keane and
  Wolpin}{1997}]{keane1997career}
---\hspace{-.1pt}---\hspace{-.1pt}--- (1997): \enquote{The career decisions of
  young men,} \emph{Journal of Political Economy}, 105, 473--522.

\bibitem[\protect\citeauthoryear{Kristensen, Mogensen, Moon, and
  Schjerning}{Kristensen et~al.}{2021}]{kristensen2021solving}
\textsc{Kristensen, D., P.~K. Mogensen, J.~M. Moon, and B.~Schjerning} (2021):
  \enquote{Solving dynamic discrete choice models using smoothing and sieve
  methods,} \emph{Journal of Econometrics}, 223, 328--360.

\bibitem[\protect\citeauthoryear{Lee and Seo}{Lee and
  Seo}{2015}]{lee2015computationally}
\textsc{Lee, J. and K.~Seo} (2015): \enquote{A computationally fast estimator
  for random coefficients logit demand models using aggregate data,} \emph{The
  RAND Journal of Economics}, 46, 86--102.

\bibitem[\protect\citeauthoryear{Luo, Perrigne, and Vuong}{Luo
  et~al.}{2018}]{luo2018structural}
\textsc{Luo, Y., I.~Perrigne, and Q.~Vuong} (2018): \enquote{Structural
  analysis of nonlinear pricing,} \emph{Journal of Political Economy}, 126,
  2523--2568.

\bibitem[\protect\citeauthoryear{Pesendorfer and Schmidt-Dengler}{Pesendorfer
  and Schmidt-Dengler}{2008}]{pesendorfer2008asymptotic}
\textsc{Pesendorfer, M. and P.~Schmidt-Dengler} (2008): \enquote{Asymptotic
  least squares estimators for dynamic games,} \emph{The Review of Economic
  Studies}, 75, 901--928.

\bibitem[\protect\citeauthoryear{Pesendorfer and Schmidt-Dengler}{Pesendorfer
  and Schmidt-Dengler}{2010}]{pesendorfer2010sequential}
---\hspace{-.1pt}---\hspace{-.1pt}--- (2010): \enquote{Sequential estimation of
  dynamic discrete games: A comment,} \emph{Econometrica}, 78, 833--842.

\bibitem[\protect\citeauthoryear{Powell, Stock, and Stoker}{Powell
  et~al.}{1989}]{powell1989semiparametric}
\textsc{Powell, J.~L., J.~H. Stock, and T.~M. Stoker} (1989):
  \enquote{Semiparametric estimation of index coefficients,}
  \emph{Econometrica: Journal of the Econometric Society}, 1403--1430.

\bibitem[\protect\citeauthoryear{Rust}{Rust}{1987}]{rust1987optimal}
\textsc{Rust, J.} (1987): \enquote{Optimal replacement of GMC bus engines: An
  empirical model of Harold Zurcher,} \emph{Econometrica}, 999--1033.

\bibitem[\protect\citeauthoryear{Shen}{Shen}{1997}]{shen1997methods}
\textsc{Shen, X.} (1997): \enquote{On methods of sieves and penalization,}
  \emph{The Annals of Statistics}, 2555--2591.

\bibitem[\protect\citeauthoryear{Shen}{Shen}{1998}]{shen1998method}
---\hspace{-.1pt}---\hspace{-.1pt}--- (1998): \enquote{On the method of
  penalization,} \emph{Statistica Sinica}, 337--357.

\bibitem[\protect\citeauthoryear{Stoker}{Stoker}{1986}]{stoker1986consistent}
\textsc{Stoker, T.~M.} (1986): \enquote{Consistent estimation of scaled
  coefficients,} \emph{Econometrica: Journal of the Econometric Society},
  1461--1481.

\bibitem[\protect\citeauthoryear{Stone}{Stone}{1980}]{stone1980optimal}
\textsc{Stone, C.~J.} (1980): \enquote{Optimal rates of convergence for
  nonparametric estimators,} \emph{The annals of Statistics}, 1348--1360.

\bibitem[\protect\citeauthoryear{Stone}{Stone}{1985}]{stone1985additive}
---\hspace{-.1pt}---\hspace{-.1pt}--- (1985): \enquote{Additive regression and
  other nonparametric models,} \emph{The annals of Statistics}, 13, 689--705.

\bibitem[\protect\citeauthoryear{Su and Judd}{Su and
  Judd}{2012}]{su2012constrained}
\textsc{Su, C.-L. and K.~L. Judd} (2012): \enquote{Constrained optimization
  approaches to estimation of structural models,} \emph{Econometrica}, 80,
  2213--2230.

\bibitem[\protect\citeauthoryear{Sweeting}{Sweeting}{2013}]{sweeting2013dynamic}
\textsc{Sweeting, A.} (2013): \enquote{Dynamic product positioning in
  differentiated product markets: The effect of fees for musical performance
  rights on the commercial radio industry,} \emph{Econometrica}, 81,
  1763--1803.

\bibitem[\protect\citeauthoryear{Van~der Vaart}{Van~der Vaart}{2000}]{van2000}
\textsc{Van~der Vaart, A.~W.} (2000): \emph{Asymptotic Statistics}, Cambridge
  University Press.

\bibitem[\protect\citeauthoryear{Van~der Vaart and Wellner}{Van~der Vaart and
  Wellner}{1996}]{van1996}
\textsc{Van~der Vaart, A.~W. and J.~A. Wellner} (1996): \emph{Weak Convergence
  and Empirical Processes with Application to Statistics}, New York, Springer.

\end{thebibliography}

\end{document}